%% file: main-draft.tex
\documentclass[aps,prb,showpacs,preprintnumbers,twocolumn,superscriptaddress]{revtex4-2}
\usepackage{amsmath,amssymb}
\usepackage{bm}
\usepackage{tipa}
\usepackage{upgreek}
\usepackage{comment}
\usepackage{mathrsfs}
\usepackage{graphicx}

\usepackage{braket}
\usepackage{enumitem}
\usepackage{mathbbol}
\usepackage{booktabs}

\usepackage{gensymb}
\usepackage[normalem]{ulem}
\usepackage{color}
\usepackage[colorlinks,bookmarks=true,citecolor=blue,linkcolor=red,urlcolor=blue]{hyperref}
\usepackage{hyperref}
\renewcommand{\vec}[1]{\mathbf{#1}}
\usepackage{pifont}

\allowdisplaybreaks

\usepackage{siunitx}
\usepackage{soul}

\begin{document}

	\title{Two-dimensional Shiba lattices as possible platform for crystalline topological superconductivity} 

	\author{Martina O.\ Soldini$^{1}$, Felix K{\"u}ster$^2$, Glenn Wagner$^1$, Souvik Das$^2$, Amal Aldarawsheh$^{3,4}$, Ronny Thomale$^{5, 6}$, Samir Lounis$^{3, 4}$, Stuart S.\ P.\ Parkin$^2$, Paolo Sessi$^2$, Titus Neupert$^1$}

	\affiliation{
	University of Zurich, Winterthurerstrasse 190, 8057 Zurich, Switzerland,\\
	$^2$Max Planck Institute of Microstructure Physics, Halle, Germany,\\
	$^3$Peter Grünberg Institut and Institute for Advanced Simulation, Forschungszentrum Jülich $\&$ JARA, Jülich, Germany.\\
    $^4$Faculty of Physics, University of Duisburg-Essen and CENIDE, Duisburg, Germany.\\
	$^5$ Institut fur Theoretische Physik and Astrophysik, Universität Würzburg, Würzburg, Germany.\\
	$^6$Department of Physics and Quantum Centers in Diamond and Emerging Materials (QuCenDiEM) Group, Indian Institute of Technology Madras, Chennai, India.\\
    $^*$ Corresponding authors: Martina O. Soldini, martina.soldini@physik.uzh.ch; Felix K{\"u}ster, fkuester@mpi-halle.mpg.de.
	}

	\begin{abstract}
Localized or propagating Majorana boundary modes are the key feature of topological superconductors. They are rare in naturally-occurring compounds, but the tailored manipulation of quantum matter offers opportunities for their realization. Specifically, lattices of Yu-Shiba-Rusinov bound states – Shiba lattices – that arise when magnetic adatoms are placed on the surface of a conventional superconductor can be used to create topological bands within the superconducting gap of the substrate. Here, using scanning tunnelling microscopy to create and probe adatom lattices with single atom precision we reveal two signatures consistent with the realization of two types of mirror symmetry protected topological superconductors. The first has edge modes as well as higher-order corner states, and the second has symmetry-protected bulk nodal points. In principle, their topological character and boundary modes should be protected by the spatial symmetries of the adatom lattice. Our results highlight the potential of Shiba lattices as a platform to design the topology and sample geometry of 2D superconductors.
	\end{abstract}
	
 	\maketitle
 
A superconducting state can be destroyed by a sufficiently large magnetic field~\cite{mo}. 
A local version of this effect can be observed at magnetic impurities in conventional $s$-wave superconductors: Yu-Shiba-Rusinov states~\cite{Yu, Shiba, Rusinov} -- or Shiba states for short -- are electronic modes localized at the adatom whose bound state energies lie within the superconducting gap. Through their spatial or energetic separation from all other excitations, which can be adjusted by variation of both substrate and adatom species~\cite{Kuster2021,SSL2019}, they are well-controlled building blocks that allow one to create structures of bound states through magnetic adatom lattices on superconducting surfaces. This way, instead of destroying superconductivity, magnetic impurities can be used to design new superconducting electronic structures within the gap of a conventional $s$-wave superconductor with sought-after unconventional properties~\cite{Yazdani,Pientka_2015,Brydon}. Previous experimental studies have in particular focused on one-dimensional (1D) chains~\cite{Yazdani2,PhysRevLett.115.197204,Ruby2017,Schneider2020,PhysRevB.104.045406,Schneider2022,doi:10.1126/sciadv.aar5251}, which, as envisioned in Kitaev’s model~\cite{Kitaev_2001}, may host Majorana bound states at their ends. Two-dimensional (2D) Shiba lattices have been studied as potential realizations of chiral superconductivity with unidirectionally propagating Majorana edge modes~\cite{2DFMShibaLattices}, and there have been experimental realizations of 2D Shiba structures~\cite{MGG2017,doi:10.1126/sciadv.aav6600,Lileroth2021}.

Kitaev’s Majorana chain and chiral superconductors are the elementary topological phases of spin-orbit coupled magnetic superconductors in 1D and 2D, respectively, according to the topological classification of electronic matter (class D in the tenfold way~\cite{Kitaev,10fold}). However, if one includes topological phases protected through spatial symmetries, one uncovers the much richer variety of crystalline and boundary-obstructed topological superconductors~\cite{TCSC_Ono,TCSC_Shiozaki,TCSC_Wang,PhysRevX.10.041014}. In 2D, they can support two types of topologically protected boundary features: Majorana edge states and higher-order corner states~\cite{doi:10.1126/science.aah6442, HOTI, PhysRevB.97.205135, PhysRevB.97.205136Eslam}. They are, for instance, protected by mirror symmetries that leave the edge or corner invariant. Beyond fully gapped superconductors, crystalline symmetries can also protect gap nodes. The resulting nodal topological superconductors may support flatband Majorana edge states~\cite{Schnyder_2015}. 

Our work explores a route towards constructing 2D topological (crystalline) superconductors, namely a bottom-up approach where a lattice system is built up atom-by-atom. Specifically, we study Shiba lattices of chromium (Cr) atoms on the surface of superconducting niobium (Nb).   Recent works about 3$d$ transition metals on Nb already established the high versatility of this material platform~\cite{Schneider2021, PhysRevB.103.235437,Schneider2022,doi:10.1073/pnas.2210589119}.  Using a scanning tunneling microscope, we (i) arrange the Cr into a desired finite-size lattice, (ii) characterize its topography, (iii) infer the magnetic structure from measurements with a spin-polarized tip, and (iv) spectroscopically characterize the electronic structure of the Shiba lattice with a superconducting tip. We consider two distinct lattice arrangements -- rectangular and rhombic -- of the 2D Cr adatoms and build a total five different structures from them. We present theoretical and experimental evidence that points towards antiferromagnetic ordering of the magnetic moments in both lattices. In four of these structures, we observe enhanced density of states near zero bias at edges or corners  — the characteristics of topological crystalline superconductors. Only one structure is void of boundary modes. Using symmetry analysis and a theoretical model, we show how the edge modes can be naturally explained as topological flat bands protected by mirror symmetry. A more differentiated picture arises for the corner modes: On the rhombic lattice, they can be modeled as arising from a nodal Shiba lattice superconductor and are thus void of topological protection. On the rectangular lattice, theoretical modeling indicates a protection through higher-order topology. Our work highlights the potential of Shiba lattices for creating crystalline topological superconductors and the need for scrutiny in interpreting spectroscopic data on them in favor of topological modes.  

\begin{figure*}
    \centering
    \includegraphics[width=1\textwidth,page=1]{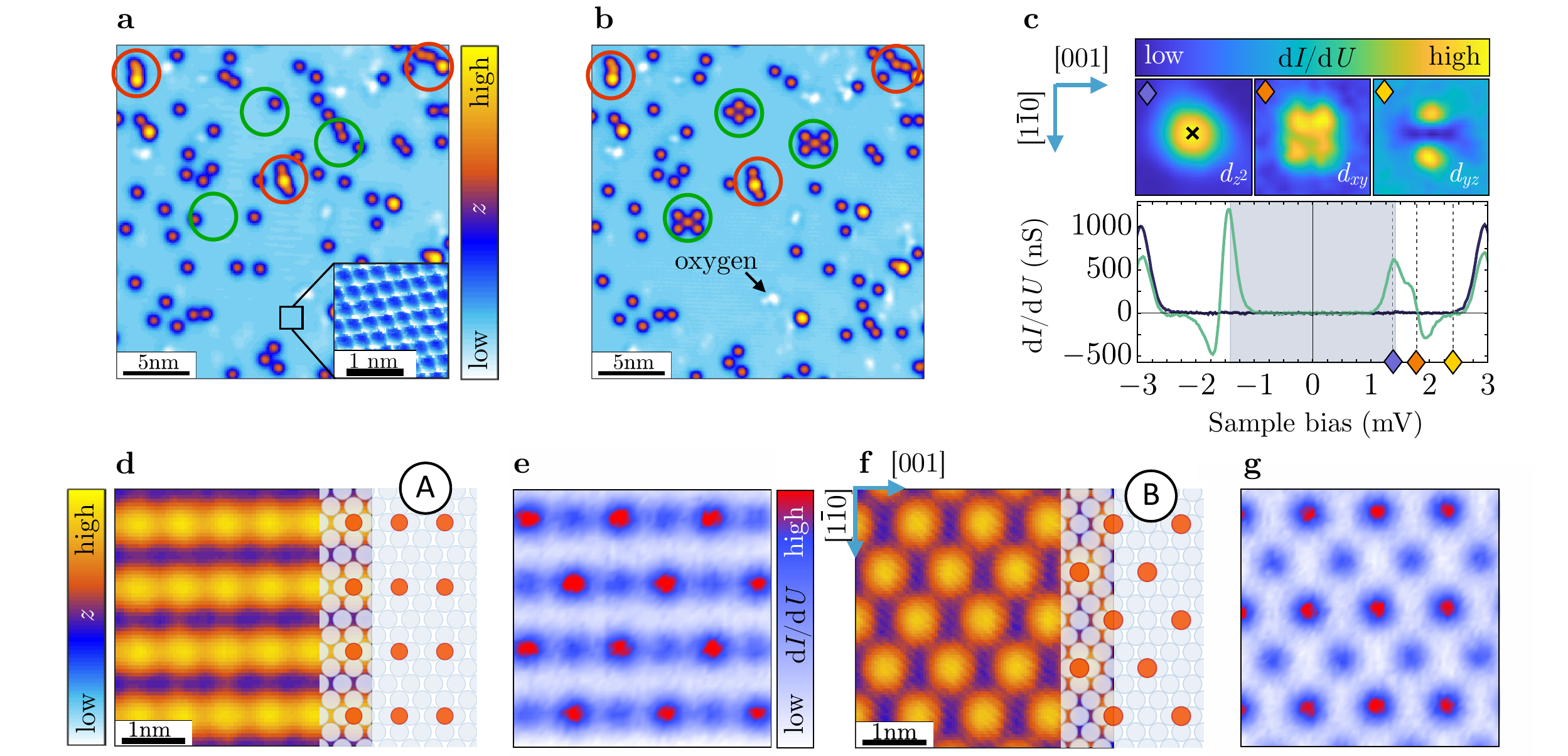}
    \caption{\textbf{System characterization}
    \textbf{a} Topographic image of the Nb(110) surface after the deposition of Cr adatoms. The inset reports an atomically resolved image of the substrate. \textbf{b} Distinct unit cells assembled by positioning single Cr adatoms using atomic manipulation techniques, with the inset showing a single Cr adatom. \textbf{c} Spectroscopy acquired on an isolated Cr adatom (green line) showing multiple Shiba bound states. A spectrum acquired by positioning the tip over the substrate is given as reference (black line). Stabilization parameters: current $500~\mathrm{pA}$, sample bias $-5~\mathrm{mV}$, modulation $50~\mathrm{\upmu V}$. The $\mathrm{d}I/\mathrm{d}U$ maps on the top display the spatial distribution of different Shiba bound states. The dominant Shiba state is of $d_{z^2}$ character. 
    \textbf{d},\textbf{f} Topographic images of two types of Cr adatoms lattices scrutinized in this work. In system $\mathsf{A}$ Cr adatoms form a rectangular lattice, while in system $\mathsf{B}$ they form a rhombic lattice. For each structure, an illustration of the position of the Cr adatoms (red dots) with respect to the underlying substrate (blue dots) is overlapped to the topographic images.
    \textbf{e},\textbf{g} spin-resolved $\mathrm{d}I/\mathrm{d}U$ maps acquired in constant-height mode on both lattices using a spin-polarized tip. In both systems, the maps reveal an alternating contrast which is indicative of an antiferromagnetic ground state. Scanning parameters: \textbf{a,b,d,f} sample bias $-5~\mathrm{mV}$, tunneling current $500~\mathrm{pA}$, inset of \textbf{a} sample bias $-5~\mathrm{mV}$, tunneling current $140~\mathrm{nA}$, \textbf{e,g} stabilization parameters $-3~\mathrm{mV}$, $500~\mathrm{pA}$, scanning bias $0.33~\mathrm{mV}$, modulation $80~\mathrm{\upmu V}$.
    }
    \label{fig:ExpPlot}
\end{figure*}

\textbf{Design of Shiba lattices.}
To create Shiba lattices, we start by depositing Cr single atoms onto the surface of a Nb(110)~\cite{PhysRevB.99.115437}. Niobium represents an optimal substrate: (i) it has the largest superconducting energy gap ($2\Delta = 3.05$~meV) among all elemental superconductors, which facilitates the spectroscopic detection of in-gap states, and (ii) it makes it possible to use atomic manipulation techniques to create nanostructures atom-by-atom using a bottom up approach~\cite{PhysRevB.102.174504,PhysRevB.103.235437,Schneider2021,Kuster2022}. This is demonstrated in Figure~\ref{fig:ExpPlot}\textcolor{red}{a}--\textcolor{red}{b}. Figure~\ref{fig:ExpPlot}\textcolor{red}{a} shows a topographic image of single Cr adatoms randomly distributed onto the clean Nb(110) substrate, with the inset reporting an atomically resolved image of the Nb(110) surface. Figure~\ref{fig:ExpPlot}\textcolor{red}{b} displays the very same sample region (red circles spotlight defects used as markers) after atomic manipulation~\cite{doi:10.1126/science.254.5036.1319} has been used to create distinct Cr nanostructures (highlighted by green circles).

Before analyzing 2D lattices, we examine the interaction between magnetic perturbations coupled to a superconducting condensate starting from the simplest case: an isolated magnetic adatom. Figure~\ref{fig:ExpPlot}\textcolor{red}{c} reports scanning tunneling spectroscopy (STS) data acquired by positioning the tip on top of an isolated Cr adatom (green line) and on the bare Nb(110) surface (black line). To enhance the energy resolution and investigate the particle-hole asymmetry of low energy modes, measurements have been acquired using a bulk Nb tip, resulting in the typical convoluted spectrum of tip and sample superconducting energy gap~\cite{doi:10.1126/science.1202204} with the gray area corresponding to the tip superconducting gap $\pm\Delta_\mathrm{tip}$ (see Extended Data Figure~\textcolor{red}{E1}). Several peaks are visible within the superconducting gap for the Cr adatom. These peaks are direct signatures of the magnetic adatom–superconductor interaction, with magnetic moments locally breaking Cooper pairs and inducing the Shiba states~\cite{Yu, Shiba, Rusinov,doi:10.1126/science.275.5307.1767}. Their distinct orbital character can be visualized by spatially mapping their wave function distribution at specific sample biases (see colored diamonds and corresponding top panels in Figure~\ref{fig:ExpPlot}\textcolor{red}{c})~\cite{PhysRevLett.117.186801,Choi2017}. The strongest intensity is observed for the $d_{z^2}$ orbital, which corresponds to the lowest-energy Shiba state located very close to $\pm\Delta_\mathrm{tip}$. Additional $d_{xy}$ and $d_{yz}$-derived Shiba states are visible at higher energies~\cite{Kuster2021,Kuster2022}. The existence of long range and highly anisotropic indirect interactions between Cr adatoms makes this platform amenable to create a large variety of distinct 2D Shiba lattices~\cite{Kuster2022}. 

Figures~\ref{fig:ExpPlot}\textcolor{red}{d},\textcolor{red}{f} show the artificial 2D lattice structures subject to the present study. For each structure, an illustration of
the position of the Cr adatoms (red dots) with respect to the underlying substrate (gray dots) is partially overlapped to the topographic
images. Figure~\ref{fig:ExpPlot}\textcolor{red}{d} (system $\mathsf{A}$) corresponds to a rectangular lattice where adatoms are placed at a next-nearest neighbor distance with respect to the underlying Nb(110) lattice. This corresponds to a periodicity of 0.66~nm along the $[001]$ direction (black line) and 0.93~nm along the $[1\overline{1}0]$ direction (green line). 
In Figure~\ref{fig:ExpPlot}\textcolor{red}{f} (system $\mathsf{B}$), Cr adatoms are arranged to create a rhombic lattice with adatoms placed at a distance of 0.85~nm along the $[1\overline{1}1]$ direction (cyan line).

The magnetic structure of the Shiba lattice crucially influences the possible (topological) features of the Shiba band structures via the symmetries it preserves/breaks. To determine the magnetic coupling between the adatoms, we performed spin-polarized measurements using a functionalized superconducting tip (see Extended Data Figure~\textcolor{red}{E2}) \cite{doi:10.1126/sciadv.abd7302}. Figures~\ref{fig:ExpPlot}\textcolor{red}{e}-\textcolor{red}{g} report spin-resolved differential conductance maps (d$I$/d$U$) acquired in constant-height mode (see Methods section) on systems $\mathsf{A}$ and $\mathsf{B}$, respectively. To stabilize the magnetic structures against fluctuations, we have performed measurements by applying an external magnetic field of 0.7~T. The d$I$/d$U$ maps show, for both lattices, an alternating spin contrast along specific crystallographic directions. In particular, an alternating contrast is visible along the $[001]$ and $[1\overline{1}0]$ directions in system $\mathsf{A}$, while it appears along the $[1\overline{1}1]$ direction in system $\mathsf{B}$. These measurements reveal the existence of an antiferromagnetic ground state for the two types of 2D lattices presented in Figure~\ref{fig:ExpPlot}\textcolor{red}{d}-\textcolor{red}{f}. In addition, we confirmed the magnetic structure with ab-initio calculations (see Methods).

\begin{figure}
    \centering
    \includegraphics[width=0.45\textwidth]{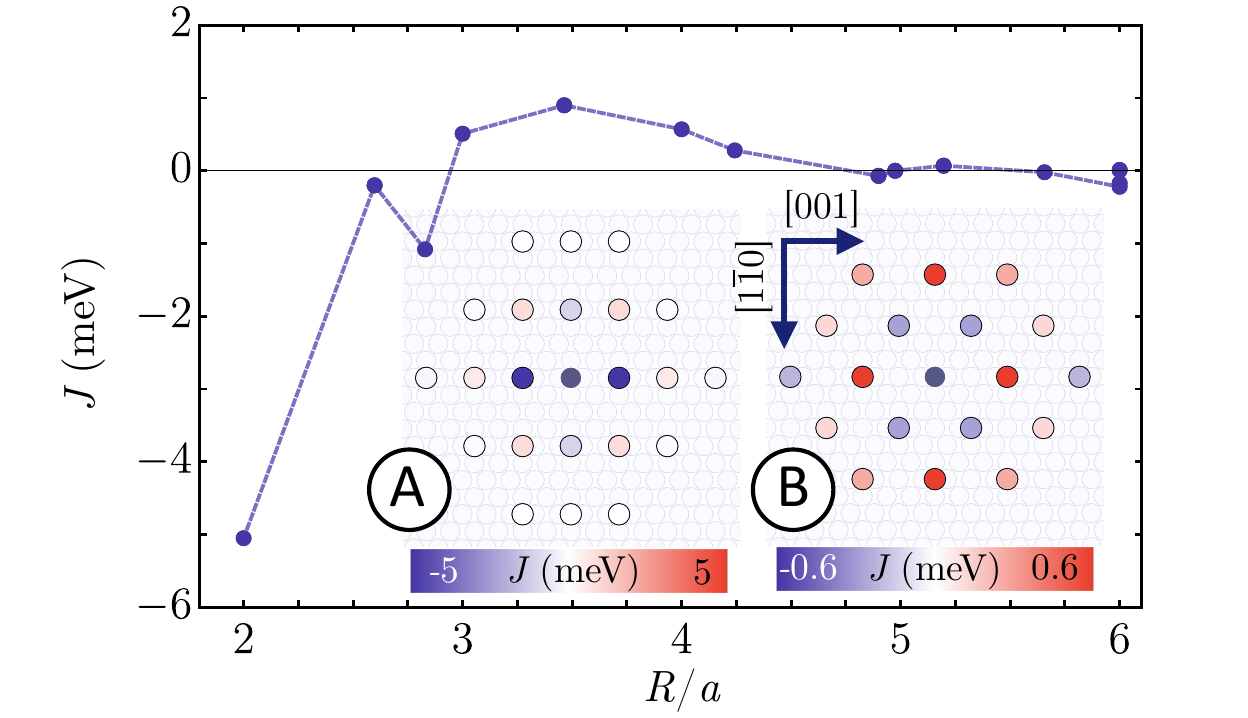}
    \caption{\textbf{Magnetic exchange interactions.}
Ab-initio Heisenberg exchange interactions $J$ as function of distance $R$ between Cr adatoms deposited on Nb(110), $a$ being the bulk Nb lattice parameter. Multiple data points at the same $R/a$ but with distinct values of $J$ correspond to symmetry non-equivalent sites. Positive (negative) values correspond to ferromagnetic (antiferromagnetic) coupling. The lower inset illustrates the simulated lattices, where each circle is colored as function of the size of $J$ with respect to the central atom (grey color). }
    \label{fig:MEI}
\end{figure}

\begin{figure*}
    \centering
    \includegraphics[width=1\textwidth]{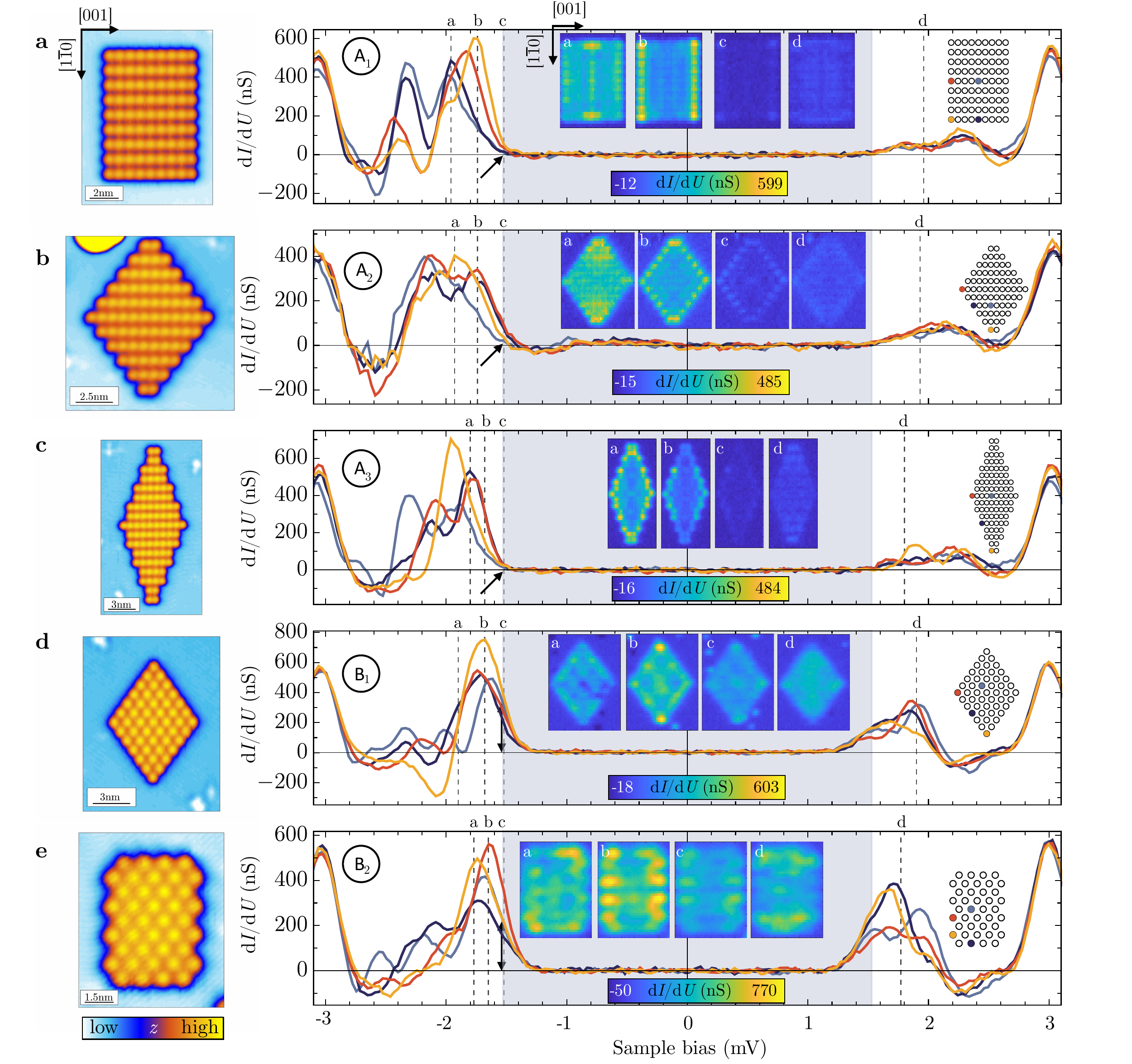}
    \caption{\textbf{Spectroscopy of Shiba lattice superconductors}
Topography of the Shiba 2D lattice termination (left) and differential conductance $\mathrm{d}I/\mathrm{d}U$ measurements (right) acquired with a superconducting tip for four representative lattices sites (see color-coded sites on the insets showing the lattice structure scheme) at the two edges, a corner, and in the bulk of each structure for     
\textbf{a} system $\mathsf{A}_1$,
\textbf{b} system $\mathsf{A}_2$,
\textbf{c} system $\mathsf{A}_3$
\textbf{d} system $\mathsf{B_1}$, and 
\textbf{e} system $\mathsf{B_2}$.
Insets show $\mathrm{d}I/\mathrm{d}U$ maps of the entire structure for selected energies marked in the main plots, labeled by the letters a--d. The color-scale within the four maps of each plot is identical, i.e., all maps referring to a specific termination have the same normalization.  The gray area corresponds to the tip superconducting gap $\pm\Delta_\mathrm{tip}$. Stabilization parameters: current $500~\mathrm{pA}$, sample bias $-5~\mathrm{mV}$, modulation $50~\mathrm{\upmu V}$.}
    \label{fig:Spectroscopy}
\end{figure*}

\textbf{Spectroscopy of Shiba lattices.}
When Cr impurities are brought close to each other, their Shiba states start overlapping and hybridizing~\cite{PhysRevLett.120.156803,Beck2021,doi:10.1073/pnas.2024837118,Kezilebieke2018,PhysRevLett.120.167001,Kuester2021,Schneider2021,LRL2022,doi:10.1073/pnas.2210589119}. 
For Cr adatoms arranged in 2D lattices, this leads to the formation of 2D Shiba bands. To scrutinize their emergence, we have acquired full spectroscopic grids for all lattices presented in the topographic images of Figures~\ref{fig:Spectroscopy}\textcolor{red}{a}--\textcolor{red}{e} (deconvoluted spectra are reported in Extended Data Figure~\textcolor{red}{E3}). Additional data demonstrating the 2D character of the Shiba bands can be found in Extended Data Figure~\textcolor{red}{E4}.  
Contrary to real materials, where crystal terminations are generally dictated by either the growth mechanism or the cleavage plane, artificial 2D lattices built atom-by-atom allow to precisely engineer their boundaries. In the following, we demonstrate how keeping the same bulk structure but specifically designing the lattice termination allows us to investigate their impact onto the emergence of distinct boundary modes.

The structures presented in Figures~\ref{fig:Spectroscopy}\textcolor{red}{a}--\textcolor{red}{c}, labeled by $\mathsf{A}_1, \mathsf{A}_2, \mathsf{A}_3$, are characterized by the same underlying lattice ordering of the Cr adatoms, corresponding to the rectangular lattice $\mathsf{A}$. The distinction between the three structures lies in the choice of different terminations: system $\mathsf{A}_1$ is terminated along the $[001]$ and $[1\overline{1}0]$ directions, system $\mathsf{A}_2$ is terminated along the $[1\overline{1}1]$ direction, and $\mathsf{A}_3$ has a double-step as an edge along $[2\overline{2}1]$.
For the systems in Figures~\ref{fig:Spectroscopy}\textcolor{red}{d},\textcolor{red}{e}, the lattice corresponds to structure $\mathsf{B}$, and is terminated along the $[1\overline{1}1]$ direction in system $\mathsf{B}_1$, while system $\mathsf{B}_2$ is terminated along two distinct and orthogonal crystallographic directions, i.e. $[001]$ and $[1\overline{1}0]$.

The spectroscopic data reported in Figures~\ref{fig:Spectroscopy}\textcolor{red}{a}--\textcolor{red}{e} compare, for systems $\mathsf{A}_1$, $\mathsf{A}_2$, $\mathsf{A}_3$, $\mathsf{B}_1$, and $\mathsf{B}_2$, d$I$/d$U$ curves obtained by positioning the tip at distinct locations, namely: corners, edges, and bulk. 
Respective atomic positions are specified with corresponding colors in insets schematically depicting the lattices. The spectra in each structure are found to be position-dependent.
In particular, the d$I$/d$U$ spectra acquired at edges and corners show stronger spectral intensity at low energies with respect to the bulk. It is noteworthy that the spectral weight appears generally particle-hole asymmetric. To visualize the spatial distribution of the local density of states (LDOS), d$I$/d$U$ maps at specific energies are reported as insets in Figures~\ref{fig:Spectroscopy}\textcolor{red}{a}--\textcolor{red}{e}. Each map is identified by a letter corresponding to a specific energy in the d$I$/d$U$ curves. A comparison of these spectroscopic data makes it possible to scrutinize the response of the electronic properties to specifically designed lattice terminations.

Structures $\mathsf{A}_1$, $\mathsf{A}_2$, and $\mathsf{A}_3$ are all built based on the rectangular lattice $\mathsf{A}$. 
On the one hand, their bulk properties appear to be termination-independent. This is highlighted in two distinct ways: (i) the d$I$/d$U$ curves acquired in the bulk (gray line) are  characterized by a similar lineshape with vanishing signal at zero energy ($\pm\Delta_\mathrm{tip}$ in our case because of the use of a superconducting tip); (ii) the spatially resolved d$I$/d$U$ maps always show vanishing intensity inside the bulk at zero energy (c insets) while the intensity becomes progressively stronger by increasing the bias (a insets), being characterized by a strong energy-dependent spatial distribution (see Extended Data Figure~\textcolor{red}{E4}) with the most prominent spectral features being of $d_{z^2}$ orbital origin (see Extended Data Figure~\textcolor{red}{E5}). These observations are consistent with Shiba bands that have a bulk gap, but our instrumental energy resolution is not high enough to prove this definitively. Also, the gap can be affected by the energy broadening associated to the quasiparticle lifetime~\cite{HPS2020} (see Extended Data Figure~\textcolor{red}{E6}).
On the other hand, the boundary modes are termination-dependent. A low energy edge mode is clearly detected in the rectangular system $\mathsf{A}_1$, as demonstrated by the d$I$/d$U$ map b in Figure~\ref{fig:Spectroscopy}\textcolor{red}{a}. The mode is localized at the boundary oriented along the $[1\overline{1}0]$ direction, while it is absent at the boundary oriented along the $[001]$ direction. For terminations $\mathsf{A}_2$ and $\mathsf{A}_3$, a boundary mode is visible around the entire edge (see insets b). Moreover, system $\mathsf{A}_3$ is characterized by an enhanced low energy LDOS close to the corners, as visible in inset b and Extended Data Figure~\textcolor{red}{E7}. These results signal how the boundary electronic properties are qualitatively distinct in structures $\mathsf{A}_1$, $\mathsf{A}_2$, and $\mathsf{A}_3$. All boundary modes are characterized by a spectral intensity primarily located on top of the adatoms. On each adatom, the spatial distribution resembles the symmetry of the single $d_{z^2}$ state. This observation allows to conclude that the boundary modes are clearly originating from the $d_{z^2}$ orbital.

In contrast to the rectangular case, the rhombic systems $\mathsf{B}_1$ and $\mathsf{B}_2$ appear gapless. This is demonstrated by the spectral intensity at the tip superconducting gap $\pm\Delta_\mathrm{tip}$, which is far from being zero for d$I$/d$U$ curves acquired in the bulk (see blue lines and arrows in Figures~\ref{fig:Spectroscopy}\textcolor{red}{d}--\textcolor{red}{e}). A gapless bulk is further confirmed by the d$I$/d$U$ maps taken at $-\Delta_\mathrm{tip}$ (see insets c in Figures~\ref{fig:Spectroscopy}\textcolor{red}{d}--\textcolor{red}{e}). Although a stronger spectral intensity is observed at the boundaries, a finite LDOS is detected also in the bulk. 
Similar to the termination dependence observed in lattice $\mathsf{A}$, systems $\mathsf{B}_1$ and $\mathsf{B}_2$ are also found to host distinct boundary modes: low energy corner modes are experimentally detected in system $\mathsf{B}_1$ (see d$I$/d$U$ map b in Figure~\ref{fig:Spectroscopy}\textcolor{red}{d}), while they are absent in system $\mathsf{B}_2$. 

Overall, these observations demonstrate that, starting from the very same building block (Cr adatoms in our case), it is possible to create both gapped (systems $\mathsf{A}_1$, $\mathsf{A}_2$, $\mathsf{A}_3$) or gapless (system $\mathsf{B}_1$ and $\mathsf{B}_2$) Shiba lattices. This goal can be  achieved  by assembling 2D nanostructures characterized by distinct symmetries. Moreover, our results reveal that, even when considering the very same 2D structure, different lattice terminations play a role in the emergence of distinct boundary modes. 

\begin{figure*}
    \centering
    \includegraphics[width=1\textwidth]{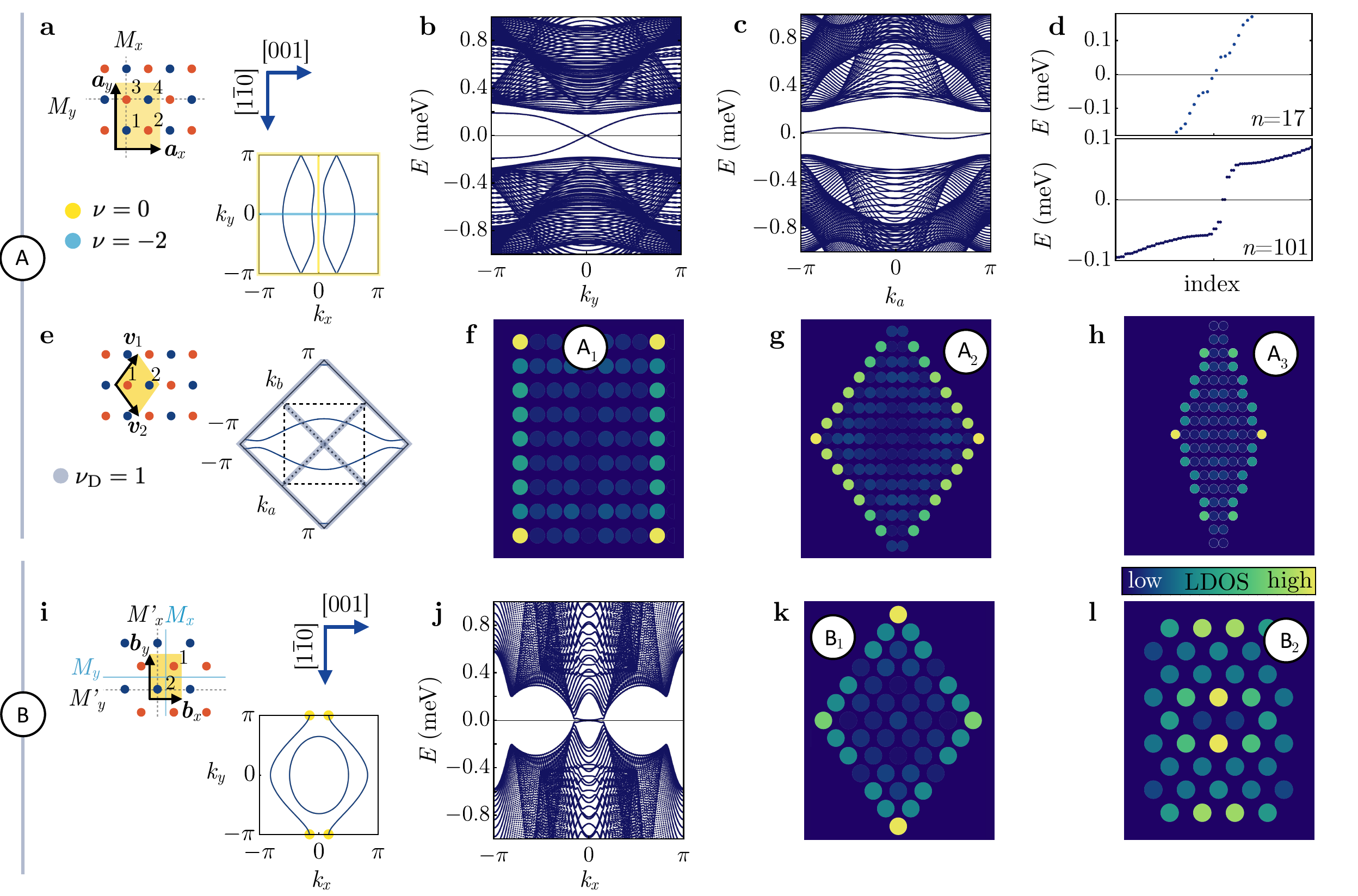}
     \caption{\textbf{Theoretical modeling of Shiba lattice superconductors.}
\textbf{a} Upper panel: Lattice structure for $\mathsf{A}$ systems, with the assumed antiferromagnetic ordering (blue: spin up, red: spin down) and lattice vectors $\vec{a}_{x/y}$. The unit cell, compatible with $\mathsf{A}_1$, is shown in yellow together with the intersection of mirror planes $M_x$ and $M_y$ with the plane of the Shiba lattice.
Lower panel: BZ of the model with normal state Fermi surfaces. The mirror invariant paths where the winding number $\nu$ is defined are marked by yellow (light-blue) lines, indicating zero (non-zero) $\nu$. \textbf{e} Analogous to \textbf{a}, with lattice vectors $\vec{v}_{1/2}$ and unit cell compatible with $\mathsf{A}_2$. Particle-hole invariant lines of the BZ are colored in gray.
\textbf{b},\textbf{c} Ribbon spectra for the $\mathsf{A}$ model with open boundary conditions along the $\bm{a}_x$, $\bm{v_1}$ direction, respectively. 
\textbf{d} Spectra for the $\mathsf{A}_3$ termination (upper panel), and for a termination of the same type but with larger size (lower panel).
\textbf{f}--\textbf{h} Theoretical electronic local density of states at zero energy in the geometry of systems \textbf{f} $\mathsf{A}_1$, \textbf{g} $\mathsf{A}_2$ and \textbf{h} $\mathsf{A}_3$ (thermal broadening: 0.1~meV).
 \textbf{i} 
Analogous to panel \textbf{a}, but for $\mathsf{B}$ systems. The lattice vectors are $\vec{b}_{x/y}$, and the intersection of glide mirror planes $M_x$ and $M_y$ and mirror planes $M_x'$ and $M_y'$ with the plane of the Shiba lattice is marked.
Yellow dots in the BZ mark the position of symmetry-protected gap nodes in the presence of superconducting pairing. 
 \textbf{j} Ribbon spectrum with open boundary conditions along the $\vec{b}_y$ direction.
 \textbf{k} and \textbf{l} same as  \textbf{f}--\textbf{h}, but for the system geometries $\mathsf{B_1}$ and $\mathsf{B_2}$, respectively.}
\label{fig:Theory}
\end{figure*}

\textbf{Theoretical Analysis.}
To elucidate a potential topological origin of the observed edge and corner modes, we introduce minimal models for the Shiba lattices.  Based on detailed symmetry considerations, we can draw conclusions about their topological properties that are robust independently of the specific choice of model parameters. We focus our models on the most pronounced and lowest energy Shiba orbital of a single Cr adatom, which is of  $d_{z^2}$ type (Figure~\ref{fig:ExpPlot}\textcolor{red}{c}). 

We developed two single-orbital tight-binding models to describe lattices composed of Shiba in-gap states. The models include up to next-to-nearest-neighbor hopping, Rashba spin-orbit coupling, the Hund's coupling between these electrons and the magnetic moment of Cr adatoms, and $s$-wave superconductivity induced from the bulk (see Methods and Supplementary Information~\textcolor{red}{II.A},~\textcolor{red}{C}). The two models are constrained by the symmetries of the rectangular and rhombic adatom lattices, respectively.
In both cases, we assume antiferromagnetic ordering of the Cr moments as suggested by experimental measurements on adatom chains, shown in the lattice schemes of Figures~\ref{fig:Theory}\textcolor{red}{a},\textcolor{red}{e},\textcolor{red}{i}, and we consider spins pointing along the out of plane ([110]) direction. In Supplementary Information~\ref{sec:Non z-aligned spins}, we comment on the consequences of relaxing the latter assumption. 

Focusing first on the rectangular lattice, the relevant spatio-temporal symmetries, which we denote by $\widetilde{M}_{x/y}$, are a composition of either of the mirror operations $M_{x/y}$ with the spinful time-reversal transformation $\mathcal{T}$, and $\widetilde{M}_{x/y}^2 = (M_{x/y}\mathcal{T})^2=+\mathbb{1}$, see Figure~\ref{fig:Theory}\textcolor{red}{a} (note that $M_{x/y}$ and $\mathcal{T}$ individually are not symmetries of the system, as they flip all spins). The paths highlighted with colors in the Brillouin Zone (BZ) of Figure~\ref{fig:Theory}\textcolor{red}{a} are left invariant under the action of particle-hole symmetry as well as  $\widetilde{M}_{x/y}$. Thus, along these 1D subspaces of momentum space, the Hamiltonian belongs to the BDI class of the tenfold way topological classification of topological insulators and superconductors~\cite{10fold}, with a $\mathbb{Z}$ topological classification. The corresponding topological invariant is the chiral winding number $\nu$. Associated with the four $\widetilde{M}_{x/y}$-invariant lines are thus four 1D winding numbers: $\nu_{y, k_x}$, for $k_x=0,\pi$ and $\nu_{x, k_y}$, for $k_y=0,\pi$. For a range of the model parameters this invariant is non-trivial ($\nu=2$) when computed along one of the lines, while the bulk is gapped. The value $\nu_{x, k_y=0,\pi}=-2$ implies a geometry-dependent bulk-boundary correspondence: 
On the one hand, a ribbon of the system with open boundary conditions (OBC) in $y$ direction will have a Dirac crossing of edge modes at zero energy and momentum $k_x=0,\pi$ along the edge (and similar for $x\leftrightarrow y$), see Figure~~\ref{fig:Theory}\textcolor{red}{b}. Specifically, for $\nu_{x,0}=2$ and $\nu_{x,\pi}=\nu_{y,0}=\nu_{y,\pi}=0$, these low energy modes are visible in the LDOS  of a finite system, as boundary modes on the edges parallel $y$, as shown in  Figure~\ref{fig:Theory}\textcolor{red}{f}. 
On the other hand, a geometry with an $\widetilde{M}_{y}$-invariant corner will support an isolated higher-order corner mode~\cite{doi:10.1126/science.aah6442,HOTI,PhysRevB.97.205135,PhysRevB.97.205136Eslam}, as the one appearing in the LDOS of the termination in Figure~\ref{fig:Theory}\textcolor{red}{h} (see Supplementary Information~\ref{subsec:HigherOrderTSC}). Two isolated zero energy modes are also visible in the open boundary condition spectra of this termination, see Figure~\ref{fig:Theory}\textcolor{red}{d}.
These predictions are consistent with the experimental observations of an increased LDOS at the $[001]$ edge of system $\mathsf{A}_1$, as shown in the inset of Figure~\ref{fig:Spectroscopy}\textcolor{red}{a}, and at the corners of system $\mathsf{A}_3$, see Figure~\ref{fig:Spectroscopy}\textcolor{red}{c}. Since the edge and corner modes are protected by crystalline lattice symmetries, the experimental terminations $\mathsf{A}_1$ and $\mathsf{A}_3$ realize a gapped topological crystalline superconductor. 

A re-definition of the unit cell, according to Figure~\ref{fig:Theory}\textcolor{red}{e}, naturally leads to the modeling of the termination $\mathsf{A}_2$. Once again, high symmetry lines in the BZ, corresponding to $k_a, k_b = 0, \pi$, are characterized by particle-hole symmetry. Hence, these 1D paths in momentum space belong to the D class of the tenfold way, with a $\mathbb{Z}_2$ classification. By a Wilson loop calculation~\cite{10fold}, we find a non-trivial topological invariant $\nu=1$ along each of these lines. This results in topological edge states (see Supplementary Information~\textcolor{red}{II.B}), visible in the ribbon spectra with open boundary conditions along $k_b$, characterised by zero energy modes at $k_a=0, \pi$ (see Figure~\ref{fig:Theory}\textcolor{red}{b}), and an enhanced LDOS at the edges of an open boundary structure (see Figure~\ref{fig:Theory}\textcolor{red}{g}). In contrast to the edge and corner modes of Figure~\ref{fig:Theory}\textcolor{red}{f},\textcolor{red}{h}, the edge modes in Figure~\ref{fig:Theory}\textcolor{red}{g} do not require crystalline symmetries to be topologically protected. Once more, this prediction is in agreement with the experimental observations of an enhanced LDOS at the edges of system $\mathsf{A}_2$ (see Figure~\ref{fig:Spectroscopy}\textcolor{red}{b}).

For the rhombic lattice, the relevant symmetries are the glide mirror symmetries $M_{x/y}$ and the composition of the two mirror operations $M'_{x/y}$ with time reversal, again denoted as $\widetilde{M}_{x/y}=M'_{x/y}\mathcal{T}$, where $\widetilde{M}_{x/y}^2=+\mathbb{1}$  (see Figure~\ref{fig:Theory}\textcolor{red}{i}). 
For this case, we find an energy structure with gapless nodal points whose degeneracy is protected by glide mirrors $M_{x/y}$ combined with particle-hole symmetry (see Supplementary Information): any crossing of the normal-state Fermi surface with the BZ boundary becomes a symmetry-protected nodal point of the gap once superconductivity is included (yellow dots in Figure~\ref{fig:Theory}\textcolor{red}{i}). Furthermore, along the paths $\bm{k}=(k_x, k_y=0)$ and $\bm{k}=(k_x=0, k_y)$ in the BZ, along which there is generically a full superconducting gap, the Hamiltonian belongs to the BDI class by virtue of $\widetilde{M}_{x/y}$ symmetry. As in the case of the rectangular lattice, a non-trivial chiral winding number along one of these paths results in a zero-energy mode in a ribbon geometry. For a specific choice of model parameters, this is shown in the ribbon spectrum (Figure~\ref{fig:Theory}\textcolor{red}{j}). However, in contrast to the case of the rectangular lattice, these edge states coexist with the bulk gapless modes. Nevertheless, they result in an enhanced LDOS at the corners of the lattice geometry $\mathsf{B}_{1}$ (Figure~\ref{fig:Theory}\textcolor{red}{k}), which agrees with the experimental observations (Figure~\ref{fig:Spectroscopy}\textcolor{red}{d}). As the bulk is gapless, this mode is not topologically protected.
For the lattice geometry $\mathsf{B}_{2}$, our model has a rather featureless LDOS at zero energy with slightly enhanced spectral weight at the edges (Figure~\ref{fig:Theory}\textcolor{red}{l}), consistent with the experimental measurements (Figure~\ref{fig:Spectroscopy}\textcolor{red}{e}). The latter example shows that the interpretation of LDOS in Shiba lattices can easily be misguided, as the corners in  Figure~\ref{fig:Theory}\textcolor{red}{k} could be misinterpreted as carrying Majorana zero modes. Nevertheless, our theoretical models for the rectangular and rhombic Shiba lattices are in topological crystalline superconducting phases with gapped and gapless 2D bulk, respectively. They demonstrate how abundant topological phases are in spin-orbit coupled Shiba lattices. 

\vspace{0.5cm}

\textbf{Discussion.}
For all the Shiba lattice structures examined in this work, our experimental data show spectroscopic signatures qualitatively consistent with the theoretical predictions of edge and higher-order corner modes. At the same time, we caution that these modes are likely not isolated Majorana states, as we have only focused on the experimentally most prominent Shiba bound state of the adatoms. The existence of multiple Shiba bands of different orbital character can significantly complicate the topological phase diagram (see Extended Data Figure~\textcolor{red}{E8})~\cite{doi:10.1073/pnas.2210589119}. To solidify our findings, it is desirable to assemble larger systems to spectroscopically map the size-dependent energy position of the boundary modes and to reach lower temperatures to disentangle their intrinsic bandwidth from thermal broadening effects. We expect that, by combining lower temperatures and large lattices, it should be possible to experimentally extract the 2D bulk band structure, as recently demonstrated in the case of 1D chains~\cite{Schneider2021}, and directly reveal the possible emergence of topologically non-trivial superconducting gaps. Another aspect motivating further investigations is the particle-hole asymmetry observed in our data.  

Overall, our approach bears the potential to realize a large variety of artificial 2D Shiba lattices in a clean and disorder-free platform and specifically enables the investigation of topological crystalline superconducting phases which remained largely unexplored in experiment so far. The combined possibilities of designing lattices characterized by different symmetries and the use of distinct magnetic elements as building blocks offer the flexibility to artificially create and control topologically protected superconducting states. This platform also represents a route to explore quantum states, topology and the interplay between symmetries and geometry of the boundaries in artificial lattices~\cite{PhotonicGraphene, TopoelCircuits, PhysRevLett.124.236404}.

For example, constructing kagome or Lieb lattices may enable the study of interaction effects in Shiba bands.

\section{Acknowledgements}
AA and SL thank M. dos Santos Dias and S. Brinker for fruitful discussions. GW acknowledges NCCR MARVEL funding from the Swiss National Science Foundation. 
TN acknowledges funding from the European Research Council (ERC) under the European Union’s Horizon 2020 research and innovation programm (ERC-StG-Neupert-757867-PARATOP). 
MOS acknowledges funding from the Swiss National Science Foundation (Project 200021E\_198011) as part of the FOR 5249 (QUAST) lead by the Deutsche Forschungsgemeinschaft (DFG, German Research Foundation). AA is funded by the the Palestinian-German Science Bridge (BMBF grant number 01DH16027) and her DFT simulations were made on the supercomputer JURECA 
at Forschungszentrum Jülich with computing time granted through JARA. RT acknowledges the Wuerzburg-Dresden Cluster of Excellence ct.qmat.

\section{Author Contributions Statement}
F.K.,  S.D., and P.S. conceived and performed the experiment. M.O.S., G.W., R.T. and T.N. have worked on the theoretical analysis. A.A. and S.L. have performed the first principles calculations. M.O.S., G.W, F.K., P.S., T.N. and S.S.P.P. wrote the manuscript with input from all the authors.

\section{Competing Interests Statement}
The authors declare no competing interests.

\bibliographystyle{unsrt}
 	\bibliography{./references.bib}

\section{Methods}
\textbf{STM and STS measurements.}
All experiments were performed in a ultra-high vacuum STM setup, operated at a temperature of 500 mK. The bulk single-crystal Nb(110) substrate was cleaned by repeatedly flashing the surface to about 2300~K.  Cr adatoms were deposited in-situ and with the substrate below a temperature of 15~K. Cr adatoms could be moved with atomic precision by approaching them with the STM tip in constant current mode and a setpoint of $-5$~mV; 70~nA. d$I$/d$U$ spectra were taken by lock-in technique.
\\

\textbf{Technical details/Method -- Ab-initio.}

To assess the magnetic states of the nanostructures, we complement the spin-resolved spectroscopic measurements with first-principles simulations. Specifically, we simulate systems $\mathsf{A}$ and $\mathsf{B}$ (Figures~\ref{fig:ExpPlot}\textcolor{red}{d},\textcolor{red}{f}) assuming periodic boundary conditions and extract the tensor of magnetic exchange interactions, more details are given in Methods. The Heisenberg exchange interaction $J$ shown in Extended Data Figure~\textcolor{red}{E9} is of antiferromagnetic nature at short adatom separation, and switches to ferromagnetic coupling depending on the distance between the Cr atoms. An oscillatory damped behavior is clearly observed. The Fermi surface of Nb is rather complex and anisotropic~\cite{Kuster2022}. Therefore, the wavelength of the oscillations is not unique and depends on the direction along which the adatoms are positioned (see inset of Extended Data Figure~\textcolor{red}{E9}). Based on the calculated magnetic interactions, we find the structures designed experimentally to be antiferromagnetic, in agreement with the spin-polarized scanning tunneling microscopy (STM) measurements performed on the chains presented in Figures~\ref{fig:ExpPlot}\textcolor{red}{e},\textcolor{red}{g},\textcolor{red}{h}. The Dzyaloshinskii-Moriya interaction is rather weak, which imposes a negligible tilting of the moments for the investigated finite structures.

The ab-initio simulations are based on the the scalar-relativistic full-electron full-potential  Korringa-Kohn-Rostoker (KKR) Green function augmented self-consistently with spin-orbit interaction~\cite{Papanikolaou:2002,Bauer:2014}.  
The local spin density approximation (LSDA) is employed for the evaluation of the exchange-correlation potential~\cite{Vosko:1980}. 
We assume an angular momentum cutoff at $\ell_{\text{max}} = 3$ for the orbital expansion of the Green function. A slab is used to model the Nb surface by considering 22 layers enclosed by two vacuum regions with a thickness of \SI{9.33}{\angstrom} each. On top of one of the Nb surfaces we place the diluted Cr layer such that the adatoms reside on the hollow stacking site relaxed towards the surface by \SI{20}{\percent} of the inter-layer distance of the underlying Nb(110) surface. This  was shown to be the energetically favored stacking of single adatoms in Ref.~\onlinecite{Kuester2021}. The magnetic exchange interactions were obtained using the magnetic force theorem in the frozen-potential approximation and the infinitesimal rotation method~\cite{Liechtenstein1987,Ebert2009}.
\\

\textbf{Tight-Binding Models}
For the description of the lattice terminations $\mathsf{A}$ and $\mathsf{B}$, we construct a minimal tight-binding model for $d_{z^2}$-like orbitals. 
For $\mathsf{A}$, the normal state Hamiltonian of the lattice, with reference to the unit cell shown in Figure~\ref{fig:Theory}\textcolor{red}{a}, is
\begin{equation}\label{eq:Norm state Hamiltonian A}
\begin{split}
    \hat{H}_{\mathrm{n}} &= \sum_{\bm{r}, \sigma}\Big\{ \sum_{l=1}^4 [J\sigma(-1)^{f(l)}-\mu] \hat{c}^{\dagger}_{\bm{r} \sigma l}
    \hat{c}_{\bm{r} \sigma l} \\
    &+\sum_{n=0,1 } \Big[\sum_{l={1, 3}} t_x (\hat{c}^{\dagger}_{\bm{r} \sigma l}
    \hat{c}_{\bm{r}+n\bm{a}_x \sigma l+1}+h.c.)\\
    &+\sum_{l=1, 2}t_y (\hat{c}^{\dagger}_{\bm{r} \sigma l}
    \hat{c}_{\bm{r}+n\bm{a}_y \sigma l+2} +h.c.)\\
    &+\sum_{\sigma', l=1, 3}\lambda_x \mathrm{i}\sigma^{y}_{\sigma \sigma'}(-1)^{n+\tilde{l}}
    (\hat{c}^{\dagger}_{\bm{r} \sigma l}
    \hat{c}_{\bm{r}+n\bm{a}_x \sigma' l+1}+h.c.)\\
    &+\sum_{\sigma', l=1, 2}\lambda_y\mathrm{i}\sigma^{x}_{\sigma \sigma'}(-1)^{n+l}(\hat{c}^{\dagger}_{\bm{r} \sigma l}
    \hat{c}_{\bm{r}+n\bm{a}_y \sigma' l+2} +h.c.)
    \Big]
    \Big\},
\end{split}
\end{equation}
with $f(l):\{1,2,3,4\}\rightarrow\{0, 1, 1, 0\}$ reproducing the antiferromagnetic pattern in the unit cell, and $\tilde{l}=1,2$ for $l=1,3$ respectively.
Here, $J$ indicates the Hund's coupling strength between electrons and Cr magnetic moments, $\mu$ the chemical potential, $t_{x/y}$ and $\lambda_{x/y}$ are hopping and spin-orbit-coupling (SOC) amplitudes, while $\sigma^i$'s indicate Pauli matrices.
The Bogoliubov de Gennes Hamiltonian of the Shiba lattice then becomes
\begin{equation}\label{eq:BdG Hamiltonian A}
    \hat{H} = \sum_{\bm{k}} \Psi_{\bm{k}}^{\dagger}\begin{pmatrix}
    H_{\mathrm{n}}(\bm{k}) & \Delta \\
    \Delta^{\dagger} & -H^*_{\mathrm{n}}(-\bm{k})
    \end{pmatrix}\Psi_{\bm{k}}
\end{equation}
with $H_{\mathrm{n}}(\bm{k})$ the Bloch Hamiltonian derived from~\eqref{eq:Norm state Hamiltonian A}, $\Delta = \mathbb{1}_{4\times4} \otimes (\mathrm{i}\sigma^y)$ the $s$-wave superconducting pairing and spinor
$\Psi_{\bm{k}} = (\hat{c}_{\bm{k} \uparrow 1}, \hat{c}_{\bm{k} \downarrow 1},...,\hat{c}_{\bm{k} \uparrow 4}, \hat{c}_{\bm{k} \downarrow 4}, \hat{c}^{\dagger}_{-\bm{k} \uparrow 1}, \hat{c}^{\dagger}_{-\bm{k} \downarrow 1},...,\hat{c}^{\dagger}_{-\bm{k} \uparrow 4}, \hat{c}^{\dagger}_{-\bm{k} \downarrow 4})$.
In the chiral symmetry ($\mathcal{C}=\mathcal{P}\widetilde{\mathcal{T}}$) basis, the Bloch Hamiltonian~\eqref{eq:BdG Hamiltonian A} takes the form
\begin{equation}\label{eq:chiral decompos H}
    H(\bm{k}) = \begin{pmatrix}
    0 & q(\bm{k}) \\
    q^{\dagger}(\bm{k}) & 0 
    \end{pmatrix},
\end{equation}
where we introduced the chiral Hamiltonian $q(\bm{k})$. The chiral winding number $\nu \in \mathbb{Z}$ of the BDI class of the tenfold way can be obtained by computing the following integral along 1D closed paths in the Brillouin zone (BZ)~\cite{10fold}
\begin{equation}\label{eq:chiral winding number}
    \nu = -\frac{\mathrm{i}}{2\pi} \int^{\pi}_{-\pi} dk \ \mathrm{Tr}[q^{\dagger}(k) \partial_k q(k)],
\end{equation}
where we parametrized the path by the variable $k\in[-\pi, \pi)$.
For example, for $\nu_{x, k_x=0}$, the relevant path in momentum space is defined by $\bm{k}=(0, k)$, with $k\in[-\pi, \pi)$.
This chiral winding number is the BDI class topological invariant mentioned in the text and highlighted in Figure~\ref{fig:Theory}\textcolor{red}{a}.

For the lattice of type $\mathsf{B}$, and with reference to the unit cell shown in Figure~\ref{fig:Theory}\textcolor{red}{e}, the normal state Hamiltonian is
\begin{equation}\label{eq:Norm state Hamiltonian B}
\begin{split}
    \hat{H}_{\mathrm{n}} =& \sum_{\bm{r}, \sigma}\Big\{ \sum_{l=1}^2 [J\sigma(-1)^l-\mu] \hat{c}^{\dagger}_{\bm{r} \sigma l}
    \hat{c}_{\bm{r} \sigma l} \\
    &+ \sum_{\bm{b}}\Big[ t  (\hat{c}^{\dagger}_{\bm{r} \sigma 1}
    \hat{c}_{\bm{r}+\bm{b} \sigma 2} + h.c.) \\
    &\quad +\sum_{\sigma'}\mathrm{i}(\bm{d}(\bm{b})\times \bm{\sigma})_{\sigma \sigma'}(\hat{c}^{\dagger}_{\bm{r} \sigma 1}
    \hat{c}_{\bm{r}+\bm{b} \sigma' 2} + h.c.)
    \Big]\\
    &+ \sum_{l=1}^ 2 \Big[ t_x(\hat{c}^{\dagger}_{\bm{r} \sigma l}
    \hat{c}_{\bm{r}+\bm{b}_x \sigma l} + h.c.) \\
    &\quad+
    t_y(\hat{c}^{\dagger}_{\bm{r} \sigma }
    \hat{c}_{\bm{r}+\bm{b}_y \sigma l} + h.c.)
    \\
    & \qquad + \sum_{\sigma'} \lambda_x (\mathrm{i}\sigma^{y}_{\sigma \sigma'} \hat{c}^{\dagger}_{\bm{r} \sigma l}
    \hat{c}_{\bm{r}+\bm{b}_x \sigma' l} + h.c.)\\
    & \qquad \quad +
    \lambda_y (\mathrm{i} \sigma^x_{\sigma\sigma'}\hat{c}^{\dagger}_{\bm{r} \sigma }
    \hat{c}_{\bm{r}+\bm{b}_y \sigma' l} + h.c.) \Big]\Big\},
\end{split}
\end{equation}
where the sum runs over $\bm{b}\in \{0,\bm{b}_x,\bm{b}_y,\bm{b}_x+\bm{b}_y\}$, $t$ and $\bm{d}(\bm{b}) = d_x(\bm{b}) \hat{\bm{x}}+d_y(\bm{b}) \hat{\bm{y}}$ are nearest-neighbor hopping and spin orbit-coupling vector, $t_{x/y}$ and $\lambda_{x/y}$ are next-to-nearest-neighbor hoppings and Rashba SOC amplitude respectively.
In the last line, we took into account next-to-nearest neighbor hoppings to lift the degeneracy of additional non-symmetry protected gapless points.
The Bogoliubov de Gennes Hamiltonian assumes the same structure of the one in Eq.~\eqref{eq:BdG Hamiltonian A}, with $H_{\mathrm{n}}(\bm{k})$ obtained from~\eqref{eq:Norm state Hamiltonian B}, $\Delta = \mathbb{1}_{2\times2} \otimes (\mathrm{i}\sigma^{y})$ and spinor $\Psi_{\bm{k}}= (\hat{c}_{\bm{k} \uparrow 1}, \hat{c}_{\bm{k} \downarrow 1},\hat{c}_{\bm{k} \uparrow 2}, \hat{c}_{\bm{k} \downarrow 2}, \hat{c}^{\dagger}_{-\bm{k}, 1, \uparrow}, \hat{c}^{\dagger}_{-\bm{k} \downarrow 1},\hat{c}^{\dagger}_{-\bm{k} \uparrow 2}, \hat{c}^{\dagger}_{-\bm{k} \downarrow 2})$.

In the Supplementary Information~\textcolor{red}{II.A},\textcolor{red}{C}, we discuss in more detail the model Hamiltonians presented here. There, the analysis of the phase diagrams for the $\mathsf{A}$ and $\mathsf{B}$ lattices in terms of the values of the various parameters is carried out, and we derive the constraints on the parameters for the topological phases to exist. With this analysis, we show that the topological phases are realized for an extended range of the parameters, which is compatible with the expectations on the values inferred from the microscopic properties of the system. Therefore, the results obtained from the tight-binding models, such as the prediction of topological boundary modes in the various terminations, mainly rely on symmetry arguments and do not require fine tuning of the model parameters.

\section{Data availability}
Experimental data underlying this work, shown in Figures~\ref{fig:ExpPlot},~\ref{fig:MEI},~\ref{fig:Spectroscopy}, and theoretical data shown in Figure~\ref{fig:Theory} are available \href{10.6084/m9.figshare.22794413}{here (DOI: 10.6084/m9.figshare.22794413)}.

\clearpage 
\newpage

\onecolumngrid
	\begin{center}
		\textbf{\large --- Supplementary Information ---}\\
		\medskip
    Martina O.\ Soldini$^1$, Felix K{\"u}ster$^2$, Glenn Wagner$^1$, Souvik Das$^2$, Amal Aldarawsheh$^{3,4}$, 
    
    Ronny Thomale$^{5, 6}$, Samir Lounis$^{3, 4}$, Stuart S.\ P.\ Parkin$^2$, Paolo Sessi$^2$, Titus Neupert$^1$.
	
	\medskip
	\textit{
	$^1$University of Zurich, Winterthurerstrasse 190, 8057 Zurich, Switzerland,\\
	$^2$Max Planck Institute of Microstructure Physics, Halle, Germany,\\
	$^3$Peter Grünberg Institut and Institute for Advanced Simulation, \\ Forschungszentrum Jülich $\&$ JARA, Jülich, Germany.\\
	$^4$Faculty of Physics, University of Duisburg-Essen and CENIDE, Duisburg, Germany.\\
     $^5$ Institut fur Theoretische Physik und Astrophysik, Universitat Wurzburg, Wurzburg, Germany.\\
	$^6$Department of Physics and Quantum Centers in Diamond and Emerging Materials (QuCenDiEM) Group, Indian Institute of Technology Madras, Chennai, India.}

	\end{center}
	
	\setcounter{section}{0}
	\setcounter{equation}{0}
	\setcounter{figure}{0}
	\setcounter{table}{0}
	\setcounter{page}{1}
	\makeatletter
	\renewcommand{\theequation}{S\arabic{equation}}
	\renewcommand{\thefigure}{S\arabic{figure}}
	\renewcommand{\bibnumfmt}[1]{[S#1]}
    
    \section{Experiment}

    \subsection{Figure 1}
      \begin{figure}[!h]
        \centering
        \includegraphics[width=0.55\textwidth]{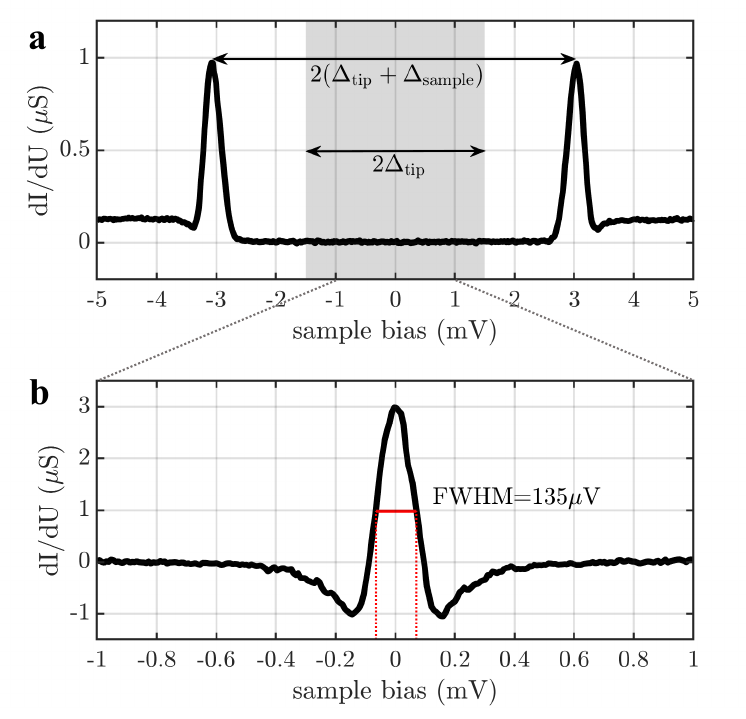}
        \caption{\textbf{Scanning tunneling spectroscopy measured with a superconducting Nb tip on clean Nb(110).} A bulk superconducting Nb cluster at the tip was obtained by deep indentation into the sample. \textbf{a} Tunneling spectroscopy shows the convolution of tip and sample density of states resulting in a large gap with size $2(\Delta_\mathrm{tip}+\Delta_\mathrm{sample})/e$. From the measured convoluted gap size of \SI{3.05}{\milli\volt} and the database value for bulk Nb ($\Delta_\mathrm{Nb}=1.52$ meV), we obtain a tip gap approximately equal to the bulk value, i.e. $\Delta_\mathrm{tip}\approxeq\Delta_\mathrm{Nb}$, which is indicated by the gray area. Measurement parameters: stabilized at sample bias \SI{-5}{\milli\volt} \textbf{b}, tunneling current \SI{500}{\pico\ampere}, bias AC modulation amplitude \SI{40}{\micro\volt}, temperature \SI{500}{\milli\kelvin}. 
        \textbf{b} Zero bias differential conductance peak. The measured FWHM of \SI{135}{\micro\volt} provides a reference for our experimental energy resolution. Measurement parameters: stabilized at sample bias \SI{-5}{\milli\volt}, tunneling current \SI{30}{\nano\ampere}, bias AC modulation amplitude \SI{10}{\micro\volt}, temperature \SI{500}{\milli\kelvin}.}
        \label{fig:EnergyResolution_SM}
    \end{figure}

\newpage
\subsection{Figure 2}
    \begin{figure}[!h]
        \centering
        \includegraphics[width=0.4\textwidth]{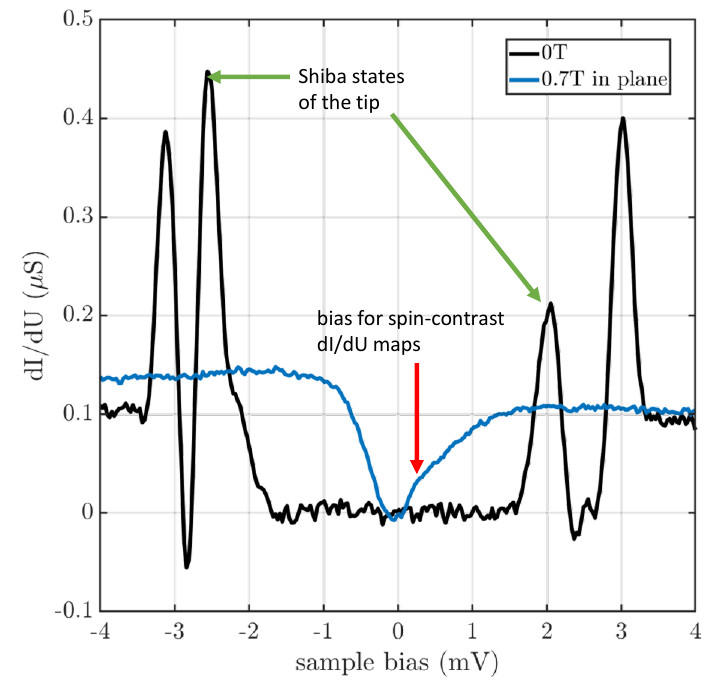}
        \caption{\textbf{Spectroscopy with spin-sensitive tip used for Figure 1e,g of the main text.} Both measurements are acquired on the clean Nb substrate with a superconducting tip featuring single Cr atoms at the apex that induce Shiba states indicated by green arrows for the zero-field measurement (black curve). The applied external magnetic field of \SI{0.7}{\tesla} in-plane overcomes the second critical field for the Nb sample while superconductivity is maintained in the Nb cluster at the tip (blue curve) where Shiba states are still visible as shoulders on the flank of the gap. The observed spin contrast in constant-height d$I$/d$U$ maps was strongest for the indicated bias at \SI{0.33}{\milli\volt}.}
        \label{fig:spinfunctip}
    \end{figure}    
\newpage
\subsection{Figure 3}
        \begin{figure}[!h]
        \centering
        \includegraphics[width=0.60\textwidth]{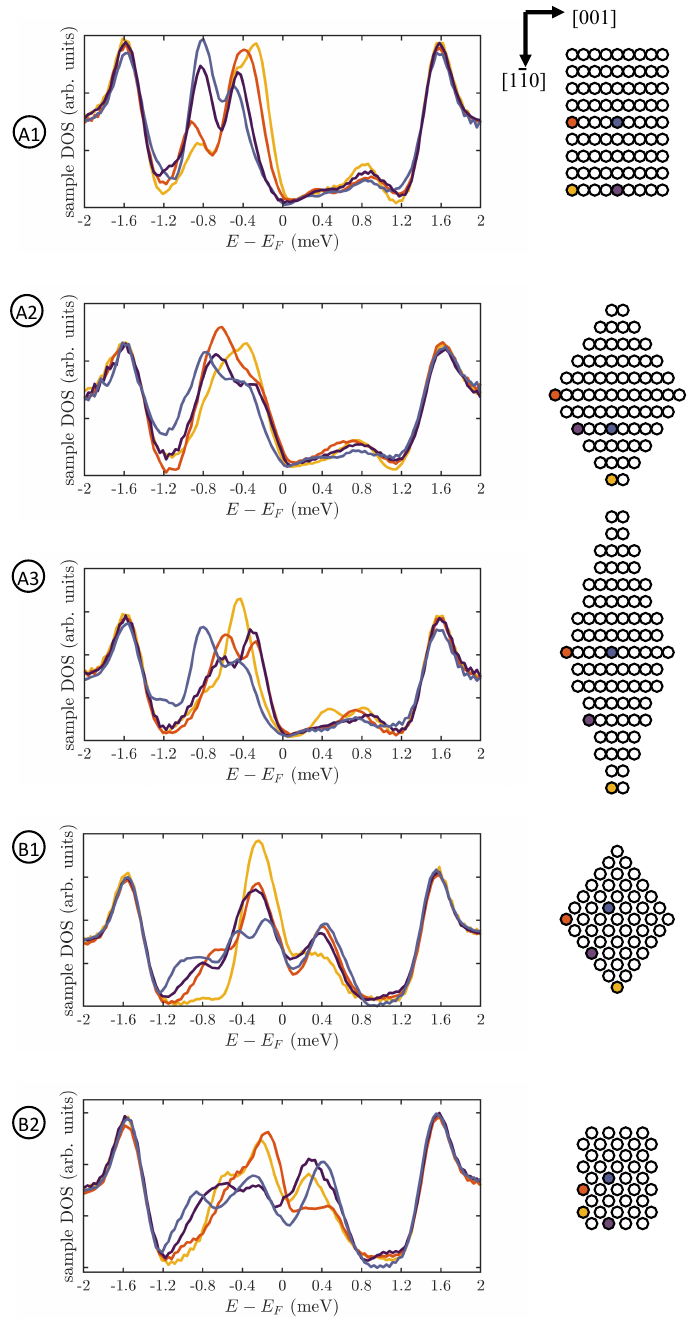}
        \caption{Tunneling spectroscopy data corresponding to figure 2 of the main article after performing numerical deconvolution. Respective lattice positions are indicated by colored circles corresponding to STS lines. Numerical deconvolution was carried out following the process described in~\cite{Schneider2021}.}
        \label{fig:Deconvoluted_SM}
\end{figure}

\newpage
    \subsection{Figure 4}
      \begin{figure}[!h]
        \centering
        \includegraphics[width=0.7\textwidth]{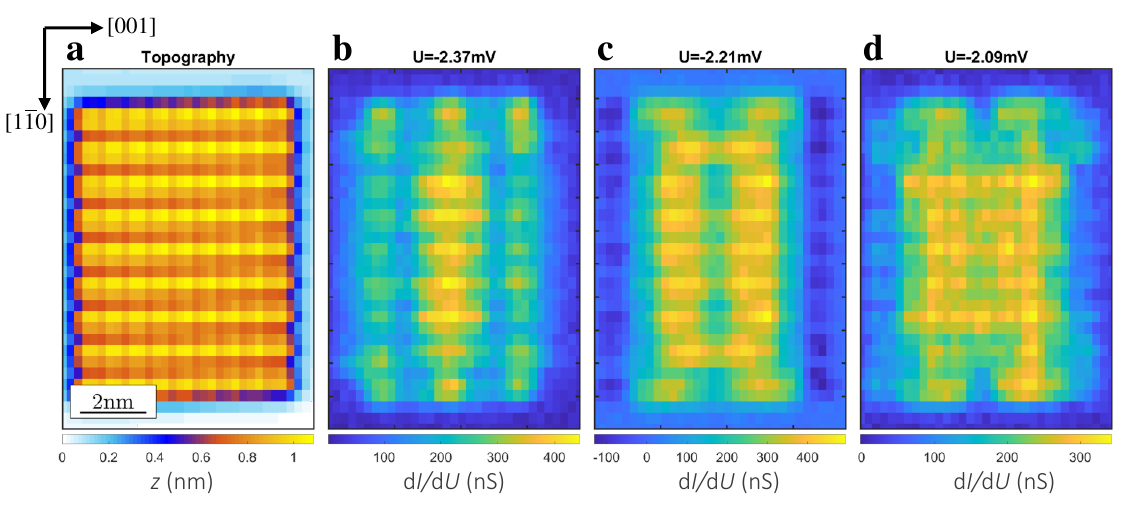}
        \caption{\textbf{Formation of two dimensional Shiba bands inside the bulk of artificial spin structures: a} Topographic image from the full spectroscopic measurement grid on structure \textsf{A$_1$}. \textbf{b,c} Two representative d$I$/d$U$ maps for bulk Shiba bands at sample biases \SI{-2.37}{\milli\volt}, \SI{-2.21}{\milli\volt} and \SI{-2.09}{\milli\volt}, demonstrating their 2D character by intensity modulations in both orthogonal directions ($[001]$ and $[1\bar{1}0]$).}
        \label{fig:BulkEnergyDependence_SM}
    \end{figure}

\newpage
\subsection{Figure 5}
  \begin{figure}[h]
    \centering
    \includegraphics[width=0.75\textwidth]{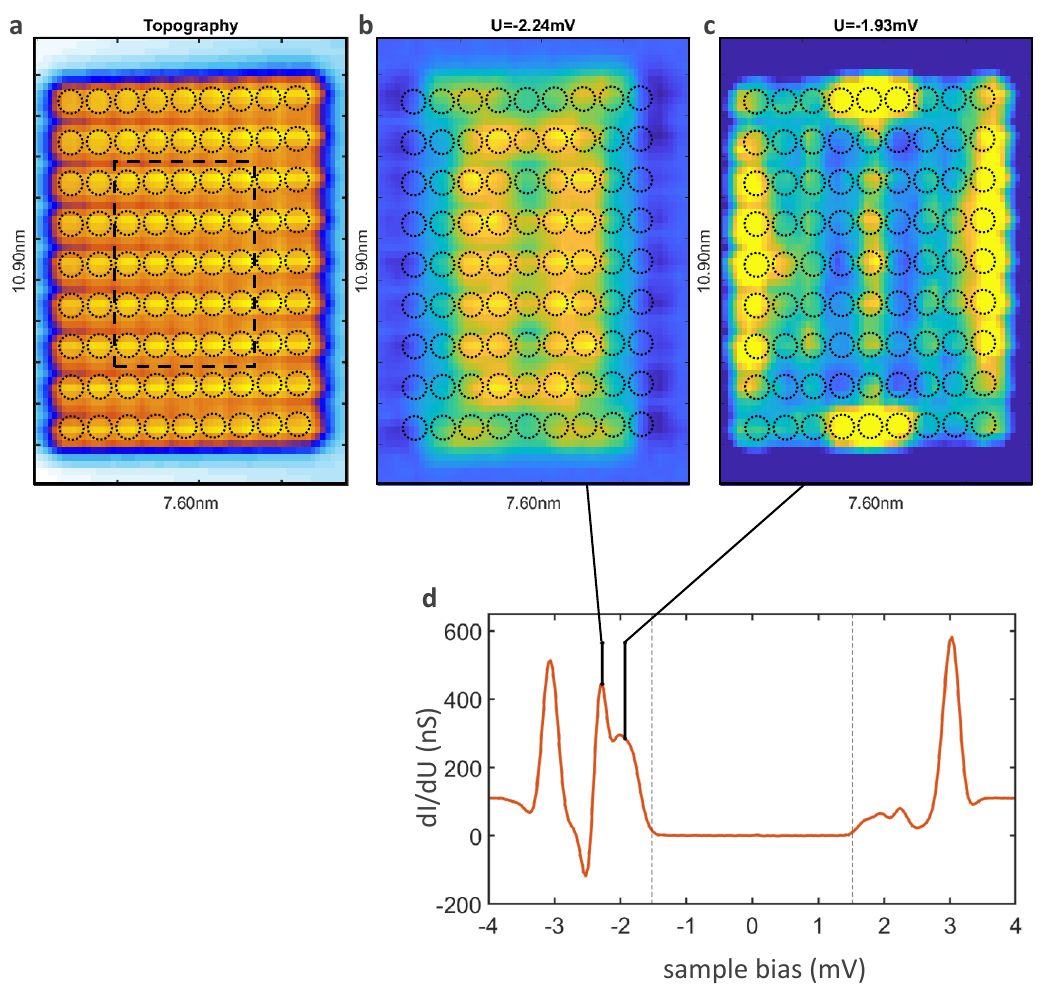}
    \caption{\textbf{Analysis of Shiba orbital characters.}\textbf{a} Topographic image of the $\mathsf{A1}$ lattice analyzed in the main text. \textbf{b-c} d$I$/d$U$ maps acquired at  biases corresponding to the two most prominent peaks in the d$I$/d$U$ spectrum reported in \textbf{d}. The spectrum has been obtained by averaging over the region identified by a dashed black box in \textbf{a}. Two peaks are dominating the scene inside the superconducting gap, centered at $U = -2.24 mV$ and $U = -2.24 mV$ (see black lines). Their strong intensity allow assigning them to $d^2_z$-derived states which can be effectively measured by STS  because of their large extension into the vacuum. This orbital assignment is corroborated by the respective  d$I$/d$U$ maps, which show two distinct patterns, both characterized by intensity maxima centered around the position of the adatoms, as expected for $d^2_z$-derived states. }
    \label{fig:OrbitalAssignment}
\end{figure}

\newpage
\subsection{Figure 6}
        \begin{figure}[!h]
        \centering
        \includegraphics[width=0.6\textwidth]{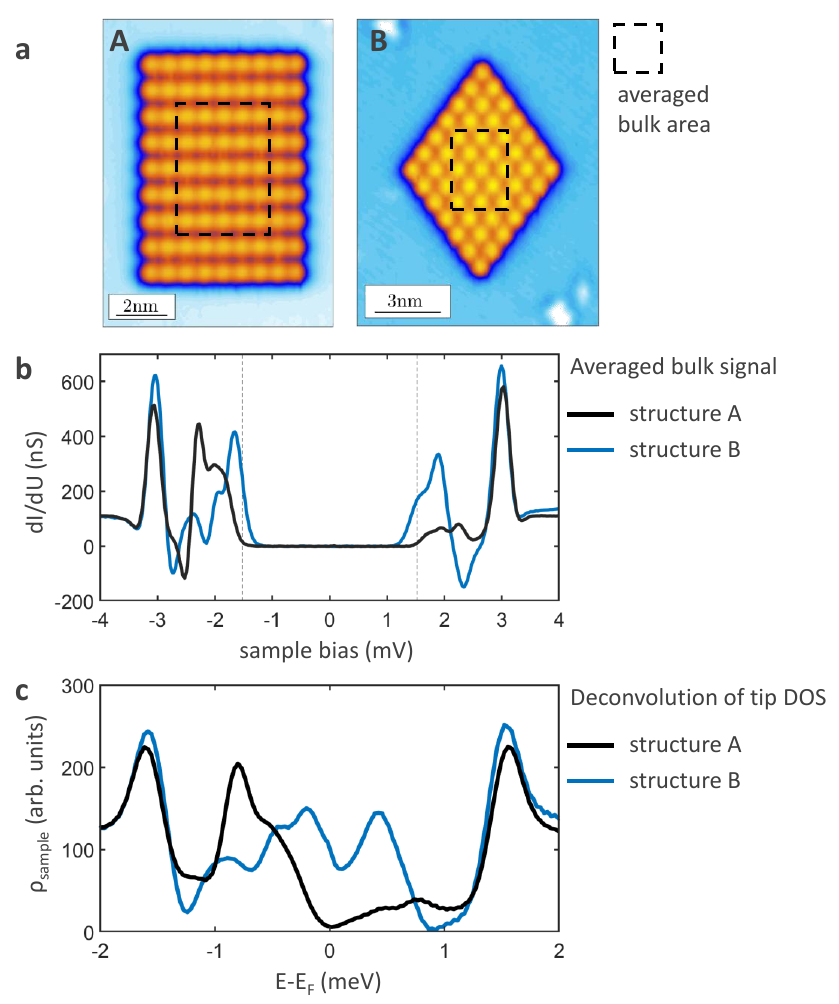}
        \caption{\textbf{Averaged bulk signals of structures A and B. a} Topography images of structures \textsf{A} and \textsf{B} with the respective bulk area indicated by a dashed rectangle used for creating a bulk STS signal which is shown in \textbf{b}. \textbf{c} plots each signal after deconvolution by a model tip DOS corresponding to a Nb superconducting gap in order to get a good approximation of the sample DOS $\rho_\mathrm{sample}$. Structure \textsf{B} has clearly no gap in the bulk while \textsf{A} features a dip at zero energy which is very close to zero intensity. 
        }
        \label{fig:averagedbulk}
    \end{figure}

\newpage

\subsection{Figure 7}
        \begin{figure}[!h]
        \centering
        \includegraphics[width=0.45\textwidth]{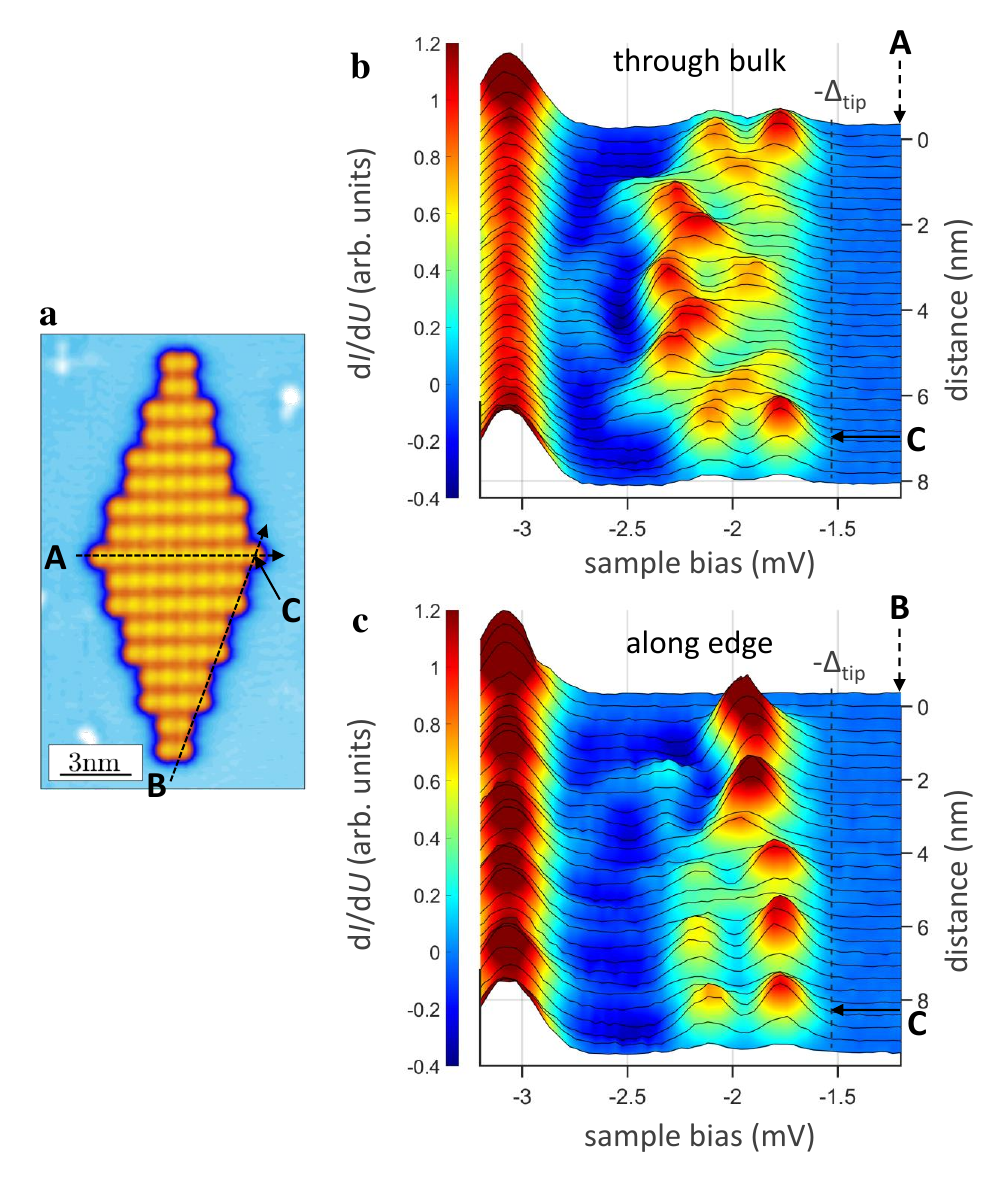}
        \caption{\textbf{Tunneling spectroscopy data acquired in several points along straight lines over the \textsf{A$_3$} termination. a} Topography image of structure \textsf{A$_3$} with two dashed lines indicating STS line paths \textsf{\textbf{A}} (through bulk) and \textsf{\textbf{B}} (along edge) as well as the corner \textsf{\textbf{C}}. \textbf{b,c} Corresponding d$I$/d$U$ spectra along those lines. At the corner \textsf{\textbf{C}}, indicated by a black solid arrow, we observe the strongest residual intensity at $-\Delta_\mathrm{tip}$ (dashed black line) corresponding to zero energy. }
        \label{fig:A3_STSlines}
    \end{figure}

\newpage
\subsection{Figure 8}
        \begin{figure}[!h]
        \centering
        \includegraphics[width=0.75\textwidth]{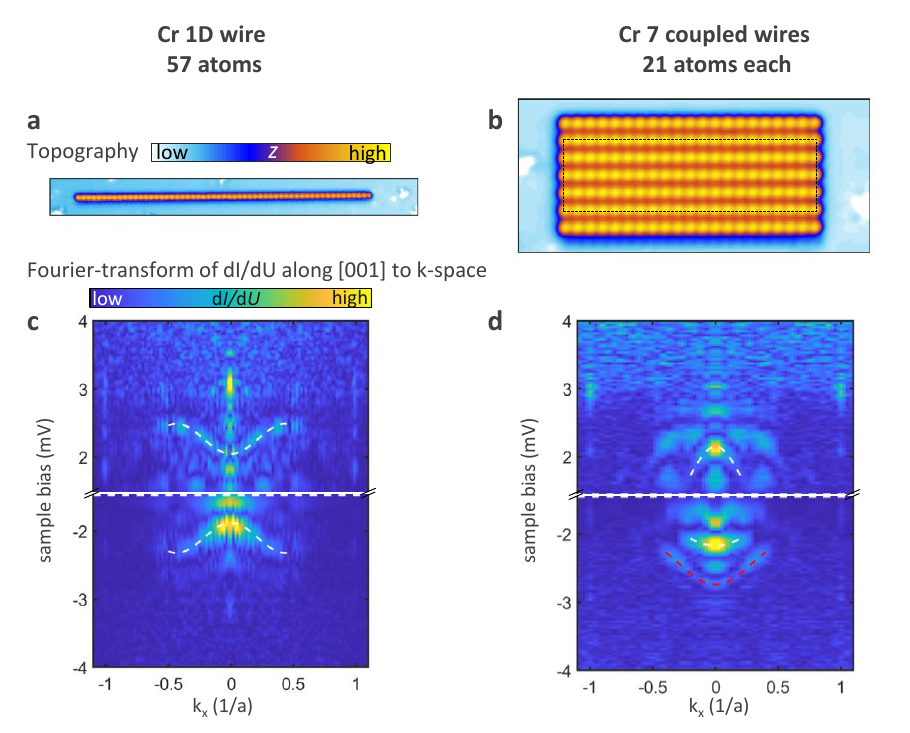}
        \caption{\textbf{Shiba bands along $[001]$ in momentum space.} Evolution of Shiba bands from 1D to 2D comparing a long 1D wire of Cr adatoms along the $[001]$ direction and \SI{0.66}{\nano\meter} inter-atomic distance to a two dimensional arrangement according to structure \textsf{A}. \textbf{a-b} Topography images.
        \textbf{c-d} FFT calculated from the STS data taken along the $[001]$ direction. For better visibility the plots are separated  to show the energy range $-4$mV...$-\Delta_\mathrm{tip}$ and $\Delta_\mathrm{tip}$...$4$mV with adjusted d$I$/d$U$ colorscale contrast. The momentum space plots suggest dispersing bands. The signal with highest intensity around $\pm2$mV is assigned to the $d_{z^2}$ atomic orbital (white dashed lines). Other bands arise from different single-atomic orbitals (red dashed line). From 1D to 2D, a change of sign in the dispersion relation is observed. This demonstrates the potential to engineer the band structure by advancement into two dimensions.
        }
        \label{fig:shibabandfft}
    \end{figure}

\clearpage
\subsection{Figure 9}
        \begin{figure}[!h]
        \centering
        \includegraphics[width=\textwidth]{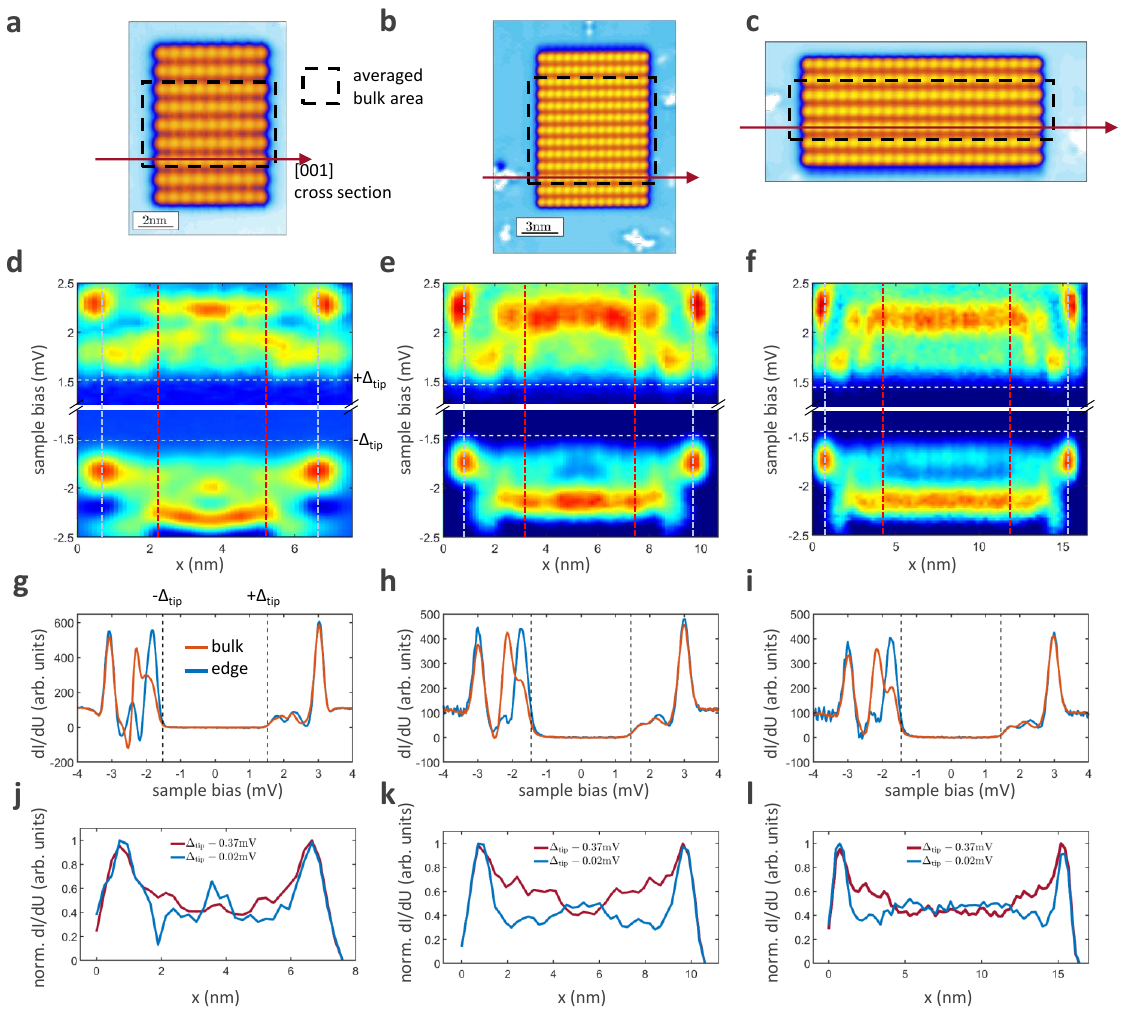}
        \caption{\textbf{d$I$/d$U$ cross sections along $[001]$. a-c} Topography images of rectangular Cr structures according to bulk structure \textsf{A} with \textsf{9x9}, \textsf{13x13} and \textsf{21x7} atoms, respectively. 
        \textbf{d-f} Cross-sectional STS along the $[001]$ direction indicated by red arrows and averaged over the bulk area shown by dashed rectangles in \textbf{a-c}. The sample bias axis is interrupted and has adjusted contrast for positive and negative bias for better visibility. Horizontal dashed lines show the tip superconducting gap, bright vertical lines indicate the positions for the edge data, red vertical lines the range of averaging for the bulk data shown in
        \textbf{g-i}. We observe equal bias positions (negative bias) for the edge state peak and the low-energy bulk peak independently from the structure size suggesting that there is a spatial extend of the edge state into the bulk which leads to non-vanishing intensity at zero energy. This is further supported by d$I$/d$U$-profiles at specific energies depicted in \textbf{j-l}. Close to zero energy the penetration depth of the edge state into the bulk is short (blue lines) compared to longer penetration at higher energies (red lines).}
        \label{fig:edgevsbulk}
    \end{figure}

\clearpage
\newpage
    \section{Theory models for the Shiba lattices}
    \subsection{Rectangular lattice}\label{subsect: SI theory A}
    In this section, we discuss the model for the rectangular lattice proposed in Eqs.~\eqref{eq:Norm state Hamiltonian A} and~\eqref{eq:BdG Hamiltonian A} of the Methods section. As described in the main text, we construct a tight binding-model for the lowest-lying Shiba orbitals, transforming in the irreducible representation of the site symmetry group of the $1a$ Wyckoff position.
    The nearest-neighbour hopping along the $[001]$ and $[1\Bar{1}0]$ crystallographic axes are $t_x$ and $t_y$ respectively. Similarly, the Rashba-spin orbit coupling strength is parametrized by $\lambda_x$ and $\lambda_y$ for the two crystallographic directions. The parameter for the Hund's coupling between the spin of the Cr magnetic moments and electron spin is $J$, and the sign in front of the interacting term alternates depending on the orientation of the adatom spin, according to the antiferromagnetic pattern.
    For convenience, we introduce the anisotropy parameter $\alpha \in [0, \pi/2]$, which interpolates between the two fully anisotropic limits ($\alpha=0, \pi/2$) and the square lattice limit ($\alpha=\pi/4$)
    \begin{equation}
        t_{x} = t \ \text{cos}\alpha, \ \ \  t_{y} = t \ \text{sin}\alpha, \ \ \  \lambda_{x} = \lambda \  \text{cos}\alpha, \ \ \  \lambda_{y} = \lambda \ \text{sin}\alpha.
    \end{equation}
    Note that for the experimental termination $\mathsf{A}$, with $[001]$ and $[1\Bar{1}0]$ corresponding to $x$ and $y$ directions, the anisotropy lies within the range $0<\alpha<\pi/2$.
    In momentum space, the normal state Bloch matrix $H_{\mathrm{n}}(\bm{k})$, in the basis corresponding to $\Psi_{\mathrm{n} \bm{k}}=(\hat{c}_{\bm{k} \uparrow 1}, \hat{c}_{\bm{k} \downarrow 1},\hat{c}_{\bm{k} \uparrow 2}, \hat{c}_{\bm{k} \downarrow 2},\hat{c}_{\bm{k} \uparrow 3}, \hat{c}_{\bm{k} \downarrow 3}, \hat{c}_{\bm{k} \uparrow 4}, \hat{c}_{\bm{k} \downarrow 4})$, reads
    \begin{equation} \label{eq:Normal state Hamilt Rect}
    H_{\mathrm{n}}(\bm{k}) = \begin{pmatrix}
        J\sigma^{z} - \mu \sigma^{0} & h_{x}(k_x) & h_{y}(k_y) & 0 \\
        h^{\dagger}_{x}(k_x) & -J\sigma^{z} - \mu \sigma^{0} & 0 & h_{y}(k_y) \\
        h^{\dagger}_{y}(k_y) & 0 & - J \sigma^{z} - \mu \sigma^{0} & h_x(k_x) \\
        0 & h^{\dagger}_y(k_y) & h^{\dagger}_x(k_x) & J \sigma^z -\mu \sigma^0
        \end{pmatrix},
    \end{equation}
    where we defined the hopping matrices as
    \begin{equation}
        h_{x}(k_x)=t_x (1+e^{\mathrm{i}k_x}) \sigma^0 +\mathrm{i} \lambda_x\sigma^y(1- e^{\mathrm{i}k_x}) \quad \text{and} \quad h_{y}(k_y)=t_y (1+e^{\mathrm{i}k_y}) \sigma^0 + \mathrm{i} \lambda_y\sigma^x(1- e^{\mathrm{i}k_y}).
    \end{equation}
    With the assumption of $J$ being the dominant energy scale, the normal state band structure splits into two sets of bands, separated by an energy gap of $\sim 2 J$.
    The Bogoliubov de Gennes Hamiltonian, with $s$-wave superconducting pairing induced from the bulk, is
    \begin{equation}\label{eq:BdG H for A - SM}
         H(\bm{k}) = \begin{pmatrix}
        H_{\mathrm{n}}(\bm{k}) & \hat\Delta \\
       \hat{\Delta}^{\dagger} & -H^{*}_{\mathrm{n}}(-\bm{k})
        \end{pmatrix}, \quad \hat{\Delta} = \Delta \begin{pmatrix}
        \mathrm{i}\sigma^y & 0 & 0 & 0 \\
          0 & \mathrm{i}\sigma^y & 0 & 0 \\
        0 & 0 & \mathrm{i}\sigma^y  & 0 \\
        0 & 0 & 0 & \mathrm{i}\sigma^y 
        \end{pmatrix},
    \end{equation}
    with basis spinor $\Psi_{\bm{k}} = (\hat{c}_{\bm{k} \uparrow 1}, \hat{c}_{\bm{k} \downarrow 1},...,\hat{c}_{\bm{k} \uparrow 4}, \hat{c}_{\bm{k} \downarrow 4}, \hat{c}^{\dagger}_{-\bm{k} \uparrow 1}, \hat{c}^{\dagger}_{-\bm{k} \downarrow 1},...,\hat{c}^{\dagger}_{-\bm{k} \uparrow 4}, \hat{c}^{\dagger}_{-\bm{k} \downarrow 4})$ and $\Delta$ the superconducting pairing amplitude.

\begin{figure}[t]
        \centering
        \includegraphics[width=0.9\textwidth]{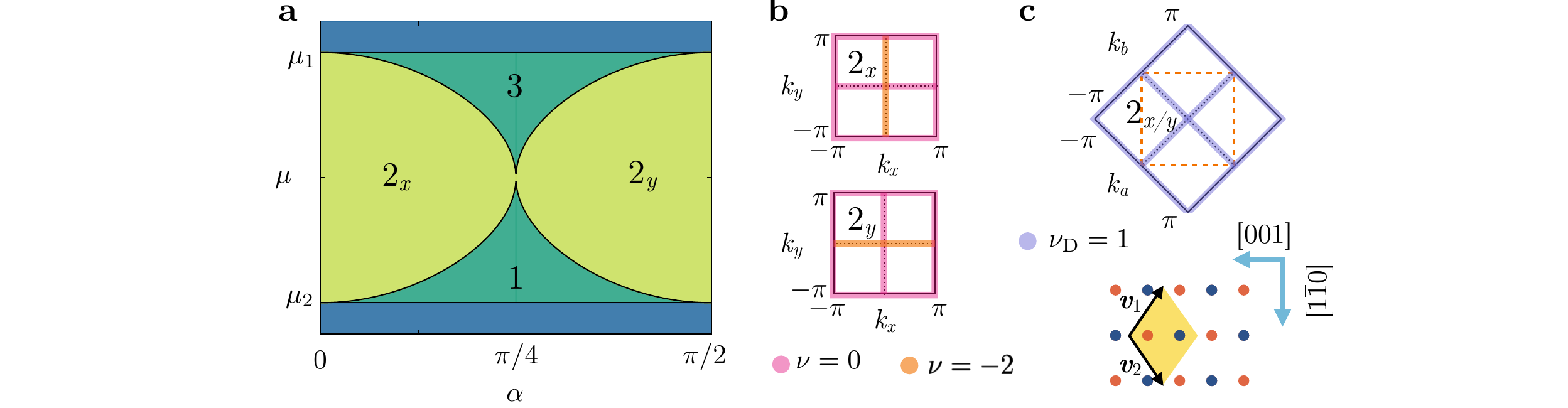}
        \caption{\textbf{a} Phase diagram for the tight binding model of Eq.~\eqref{eq:BdG Hamiltonian A}, corresponding to the rectangular lattice. The phase filled in blue is a trivial gapped phase, phases 1 and 3 are gapless, and phases $2_x$ and $2_y$ are gapped topological phases. \textbf{b} Brillouin zone of the rectangular lattice in phase $2_x$ and $2_y$. Colored lines indicate paths in $k$-space where the BDI chiral topological invariant is well-defined, and the value of the invariant ($\nu=0$, $\nu=-2$) is distinguished by the color, as listed in the legend.
        \textbf{c} Brillouin zone obtained for the alternative choice of unit cell (shown in the lower \textbf{c} panel). The lines marked in purple are those along which the superconductor belongs to the D class of the tenfold way. There, the D class topological invariant $\nu_{\mathrm{D}}$ is well-defined, and the value $\nu_{\mathrm{D}}=1$ holds in both the $2_x$ and $2_y$ phases. The dashed orange square marks the folded BZ, corresponding to the one of \textbf{b}, with the choice of unit cell considered in the main text.}
        \label{fig:PhaseDiagramA}
    \end{figure}

    The phase diagram as a function of the chemical potential $\mu$ and $\alpha$ is shown in Figure~\ref{fig:PhaseDiagramA}\textcolor{red}{a} and consists of four phases, either gapped or gapless. While the outer regions correspond to a trivial gapped phase, between the two critical values $\mu_{1,2}$ there are two gapless phases ($1$ and $3$), where four gapless Dirac cones are present, and two gapped topological phases ($2_x$ and $2_x$).
    The phase boundaries between the trivial gapped phase -unlabeled- and the two gapless phases, labeled by $1$ and $3$, are described by the $\alpha$-independent values
    \begin{equation}\label{eq:phase boundary mu 1 2}
    |\mu_{1, 2}|= \sqrt{(J\pm 2\lambda)^2-\Delta^2}
    \end{equation}
    with either positive or negative $\mu$. The symmetry under $\lambda\to-\lambda$ originates from the spinless time-reversal symmetry of the Hamiltonian
    \begin{equation}\label{eq:spinless trs}
    H(-k_x, -k_y,-\lambda_x,\lambda_y)=H(k_x, k_y,\lambda_x,\lambda_y)^*.
\end{equation}
    The separation lines between phases 1 and $2_{x/y}$ and 3 and $2_{x/y}$ correspond to the values of the chemical potential
    \begin{equation}\label{eq:phase boundary 3 and 2x}
    \begin{split}
        &|\mu_c| = \sqrt{(J\pm 2 \lambda \sqrt{|\cos(2\alpha)|})^2 - \Delta^2}, \quad \alpha \in \left[0, \frac{\pi}{2}\right],
    \end{split}
    \end{equation}
    At these critical value of the chemical potential, the bulk spectra has a gapless point along the line  $\bm{k}=(\pm k, k)$ (with the sign dependent on whether the phase is $2_{x}$ or $2_y$), at the value of $k$ that solves the equation $\text{cos}(k)=\text{cot}(\alpha)$. 
    The analytical boundaries of Eqs.~\eqref{eq:phase boundary mu 1 2} and~\eqref{eq:phase boundary 3 and 2x} constrain the values of the Hund's coupling $J$ for which the topological phase exists. From these, we get the following conditions
    \begin{equation}\
        \text{(i)} \quad J > \Delta \pm 2\lambda, \qquad \text{(ii)} \quad J > \Delta \pm 2\lambda \sqrt{|\cos(2\alpha)|},
    \end{equation}
    where (i) determines whether the gapless phases 1 and 3 exist, and (ii) is the condition for the existence of the topological phases $2_{x/y}$ at a specific value of $\alpha$.
     
    For the rectangular lattice structure, the spinful time reversal operation can be written as
    \begin{equation}
        \mathcal{T} = \begin{pmatrix}
        0 & \mathrm{i}\sigma^y & 0 & 0 \\
        \mathrm{i}\sigma^y & 0 & 0 & 0 \\
        0 & 0 & 0 & \mathrm{i}\sigma^y \\
        0 & 0 & \mathrm{i}\sigma^y & 0 \\
        \end{pmatrix} \mathcal{K}, \quad \text{or} \quad \mathcal{T} = \begin{pmatrix}
        0 & 0 & \mathrm{i}\sigma^y & 0 \\
        0 & 0 & 0 & \mathrm{i}\sigma^y  \\
        \mathrm{i}\sigma^y  & 0 & 0 &0 \\
        0 & \mathrm{i}\sigma^y  & 0 & 0 \\
        \end{pmatrix} \mathcal{K},
    \end{equation}
    where $\mathcal{K}$ indicates complex conjugation.
    With the assumption of antiferromagnetic ordering of the adatom spins, the rectangular lattice structure is characterized by two mirror operations $M_{x/y}$, which composed with time reversal lead to two spatio-temporal symmetries of the system, $\widetilde{M}_{x/y}$, as described in the main text.
    In the basis of the normal state Hamiltonian, the unitary part (i.\,e. omitting $\mathcal{K}$) of these composite symmetries can be written as
    \begin{equation}
        \widetilde{M}_{x, \mathrm{n}}(k_x) = \begin{pmatrix}
        - \mathrm{i} \sigma^z & 0 & 0 & 0 \\
        0 & - \mathrm{i} \sigma^z e^{-\mathrm{i}k_x} & 0 & 0 \\
        0 & 0 & - \mathrm{i} \sigma^z & 0 \\
        0 & 0 & 0 & - \mathrm{i} \sigma^z e^{-\mathrm{i}k_x}
        \end{pmatrix}, \qquad 
        \widetilde{M}_{y, \mathrm{n}}(k_y) = \begin{pmatrix}
        -\sigma^0 & 0 & 0 & 0 \\
        0 &  -\sigma^0 & 0 & 0 \\
        0 & 0 &  - \sigma^0 e^{-\mathrm{i}k_y} & 0 \\
        0 & 0 & 0 & - \sigma^0 e^{-\mathrm{i}k_y}
        \end{pmatrix}.
    \end{equation}
    In Nambu space, the latter two operations and particle-hole symmetry read
    \begin{align}
            \widetilde{M}_{x}(k_x) &=\begin{pmatrix}
            \widetilde{M}_{x, \mathrm{n}}(k_x) & 0 \\
            0 & \widetilde{M}^*_{x, \mathrm{n}}(-k_x)
            \end{pmatrix}, & \widetilde{M}_{x}(k_x) H(k_x, k_y)^* \widetilde{M}_{x}(k_x)^{-1} &= H(k_x, -k_y)\\
        \widetilde{M}_{y}(k_y) &=\begin{pmatrix}
            \widetilde{M}_{y, \mathrm{n}}(k_y) & 0 \\
            0 & \widetilde{M}^*_{y, \mathrm{n}}(-k_y)
            \end{pmatrix}, & \widetilde{M}_{y}(k_y) H(k_x, k_y)^* \widetilde{M}_{y}(k_y)^{-1} &= H(-k_x, k_y)\\
        \mathcal{P} &= \begin{pmatrix}
            0 & \mathbb{1}_{8 \times 8} \\
            \mathbb{1}_{8 \times 8} & 0
            \end{pmatrix}, &        \mathcal{P} H(k_x, k_y)^* \mathcal{P} ^{-1} &=- H(-k_x, -k_y).
    \end{align}
    The combination of either of the $\widetilde{M}_{x/y}$ with particle-hole symmetry defines an effective chiral symmetry $\widetilde{\mathcal{C}}_{x/y}$
    \begin{equation}
        \widetilde{\mathcal{C}}_{x/y} = \mathcal{P} \widetilde{M}_{x/y}, \qquad \widetilde{\mathcal{C}}_x H(k_x, k_y) \widetilde{\mathcal{C}}_x^{-1} = -H(-k_x, k_y), \quad \widetilde{\mathcal{C}}_y H(k_x, k_y) \widetilde{\mathcal{C}}_y^{-1} = -H(k_x, -k_y).
    \end{equation}
    As discussed in the Methods section, the Hamiltonian can be written in the basis of the operator $\widetilde{\mathcal{C}}_{x/y}$, where it assumes the chiral form of Eq.~\eqref{eq:chiral decompos H}. This in turns allows to compute the chiral winding numbers $\nu_{x, k_x}$ and $\nu_{y, k_y}$ introduced in Eq.~\eqref{eq:chiral winding number} in the Methods section.
    
    Note that the unit cell choice depicted in Figure~\ref{fig:Theory}\textcolor{red}{a} can be replaced in principle by a two-atom unit cell, as in Figure~\ref{fig:PhaseDiagramA}\textcolor{red}{c}. This naturally leads to the modeling of the $\mathsf{A}_2$ experimental termination. The BZ defined by this choice is shown in the upper panel of Figure~\ref{fig:PhaseDiagramA}\textcolor{red}{c}. The particle-hole invariant paths belong to the D class of the tenfold way, and the calculation of the Wilson loop along the paths gives a non-trivial topological invariant $\nu=1$ along any of the lines and in both phases $2_x$ and $2_y$.
    The Wilson loop is computed as~\cite{10fold}
    \begin{equation}\label{eq:Wilson Loop}
        W_{m, n}[l] = \overline{\exp}\left( \mathrm{i} \int d\bm{l} \cdot \bra{u_{m}(\bm{k})} \bm{\nabla}_{k} \ket{u_{n}(\bm{k})}\right),
    \end{equation}
    where the line over the exponent indicates the path ordering operation, $u_m(\bm{k})$ are Bloch eigenstates and $l$ is the path in momentum space. The bands considered, labeled by the indexes $m, n$, are the two bands lying closest to zero energy.
    The D class invariant is obtained as $\nu_{\mathrm{D}}[l]= \text{Arg}(\text{Tr}(W[l]))/\pi$.
    The non-trivial value of $\nu_{\mathrm{D}}$ results into zero-energy modes at $k_a=0, \pi$, visible in the ribbon spectra of Figure~\ref{fig:Theory}\textcolor{red}{c}, and analogous ones appear at $k_b=0, \pi$ in the ribbon spectra with momentum $k_b$. The OBC termination that naturally arises from opening the boundaries along the directions defined by $k_a$ and $k_b$ has zero-energy topological edge modes along each of the four edges, in both the $2_x$ and $2_y$ phases, see Figure~\ref{fig:CornerMode}\textcolor{red}{a}.
    
    For the numerical evaluation of the model, irrespective of the choice of unit cell, we choose the parameters $J=17$~meV, $\Delta=1.5$~meV, $\alpha=0.4$, $t=4$~meV, $\lambda=0.9$~meV, $\mu=17.2$~meV and thermal broadening $\varepsilon=0.1$~meV. These values are kept the same for every numerical plot referring to the rectangular lattice, including those in the main text.

\begin{figure}
        \centering
        \includegraphics[width=\textwidth]{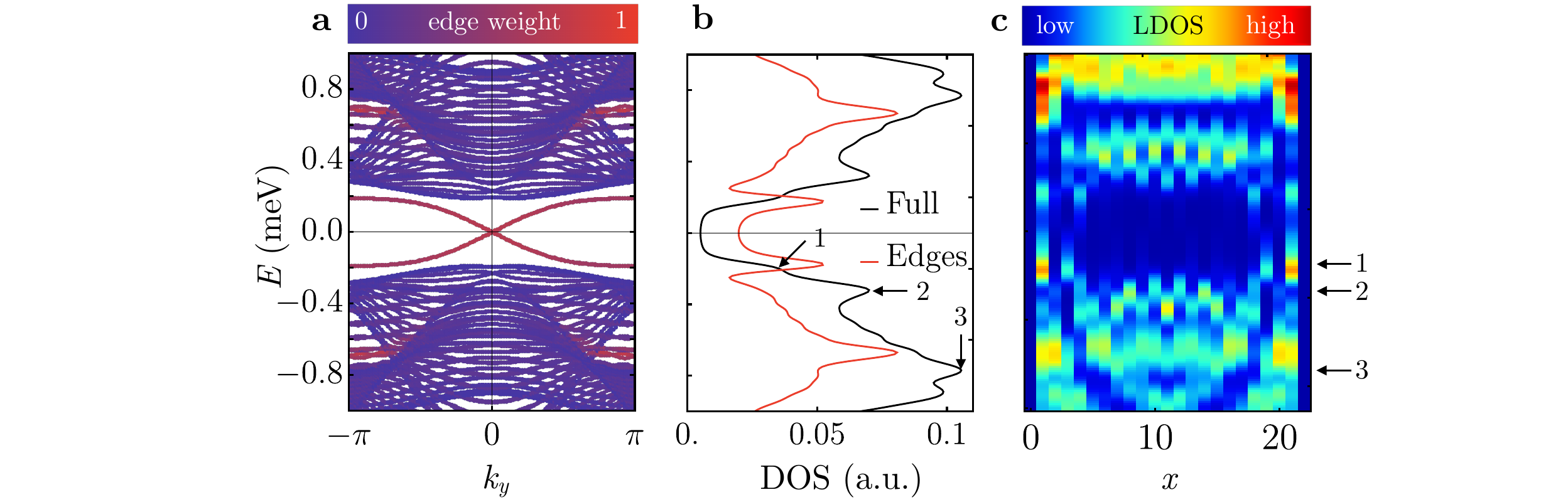}
        \caption{\textbf{Ribbon geometry with open boundary conditions along \textit{x}.} \textbf{a} Ribbon spectrum for the $\mathsf{A}$-type termination with open boundary conditions along the $x$ direction, analogous to the one presented in Fig.~\ref{fig:Theory}\textcolor{red}{b} in the main text. Here, the eigenvalues of the ribbon Hamiltonian are colored according to their eigenstate weight on the edges, i.\,e., $\sum_{\vec{r}\in \text{edges}} |\psi(\vec{r})|^2$, showing that the in-gap states are localized at the boundaries of the ribbon. \textbf{b} Density of states obtained by averaging over the ribbon momentum $k_y$, and considering a thermal broadening of $\varepsilon=0.04$ meV. The contribution to the density of states of the two unit cells at the boundaries of the ribbon is marked by the red line. The label $1$ indicates the Van Hove singularity stemming from the edge modes dispersion, while the labels $2,3$ indicate the ones arising from the bulk energy structure. The density of states of the edges is re-scaled as compared to the full DOS for visibility. \textbf{c} The LDOS for a ribbon geometry with $L_x=21$ Cr atoms along the $x$ direction is shown, as a function of position $x$ and energy. The thermal broadening is $\varepsilon=0.05$ meV. The energies corresponding to the peaks in \textbf{b} are marked by the labels $1, 2, 3$. The van Hove singularity marked by $1$ is localized at the edges of the system. Note that the colorscales in \textbf{c} is only qualitatively similar to the one in Extended Data Figure~\textcolor{red}{E6}. The data has an upper cutoff in the maximal LDOS displayed to improve the visibility.}
        \label{fig:RibbonDOS_SI}
    \end{figure}

    To provide a further comparison with the experimental data, Fig.~\ref{fig:RibbonDOS_SI}\textcolor{red}{a} shows the theoretically predicted weight of the energy modes on the edges in the ribbon geometry with open boundary conditions along $x$, alongside the density of states originating from the energy spectrum (Fig.~\ref{fig:RibbonDOS_SI}\textcolor{red}{b}). This figure shows that the peaks in the bulk LDOS appearing in the experimental measurements of Fig.~\ref{fig:Spectroscopy}\textcolor{red}{a}--\textcolor{red}{c} can be theoretically explained within a single-band model.
    In addition, Fig.~\ref{fig:RibbonDOS_SI}\textcolor{red}{c} shows the predicted LDOS of the ribbon as a function of position $x$ and energy, averaged over the momentum $k_y$. 
    
    \subsection{Higher order topological superconductor in the rectangular lattice}\label{subsec:HigherOrderTSC}
\begin{figure}[t]
        \centering
        \includegraphics[width=1\textwidth]{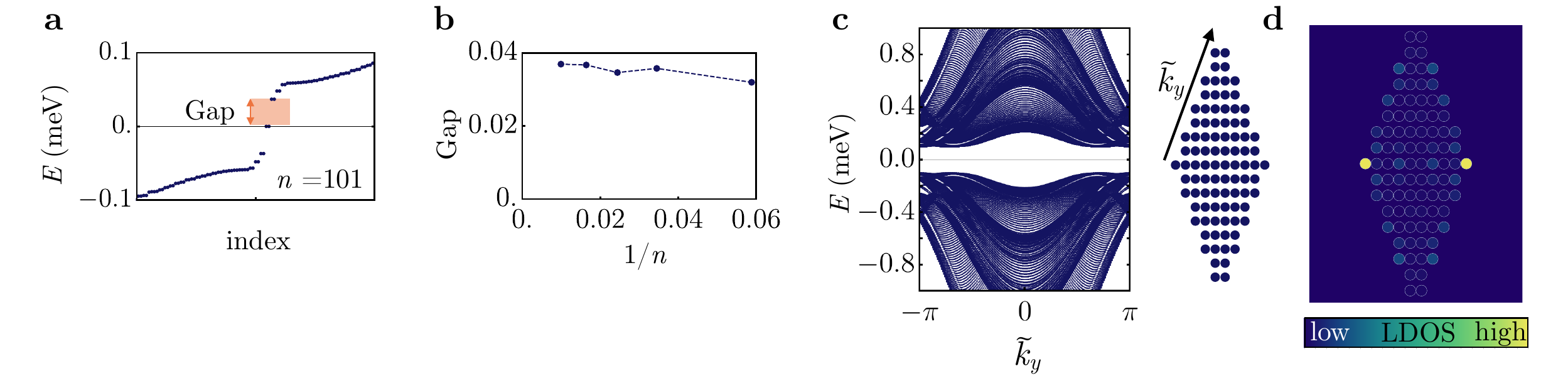}        
        \caption{\textbf{Higher order topological superconductor with corner mode} \textbf{a} Open boundary conditions spectrum for a termination of the $\mathsf{A}_3$ type, with $n=101$ ($n$ the number of horizontal rows). \textbf{b} Scaling of the gap between the two zero corner modes and the remaining edge and bulk states for $\mathsf{A}_3$-like terminations as a function of the inverse size $1/n$. \textbf{c} Ribbon calculation for the termination $\mathsf{A}_3$, with periodic direction $\tilde{k}_y$, marked in the scheme on the right. From the ribbon spectrum one sees that the mirror symmetry breaking edge is gapped. \textbf{d} Electronic part of the LDOS for the two states closest to zero energy for the termination corresponding to the experimental termination $\mathsf{A}_3$. An increased LDOS at the two corners preserving $\widetilde{M}_{x}$ is visible.}
        \label{fig:CornerMode}
    \end{figure}

    Starting from the knowledge of the crystalline symmetry protected topological nature of phases $2_x$ and $2_y$, we can design a termination where higher order topological modes appear. 
    In the geometry corresponding to the lattice realization $\mathsf{A}_1$, also reproduced theoretically in Figure~\ref{fig:Theory}\textcolor{red}{f}, each point along the two $[1\Bar{1}0]$ edges is left invariant under the action of the symmetry $\widetilde{M}_{y}$. To construct a higher order topological mode, we bend the straight edges to form an angle, such that the only boundary point left invariant under the symmetry remains a single corner, one for each one of the edges. This leads to higher order topological Majorana modes, localized at the corners and appearing at zero energy. With an appropriate choice of angle, the edge modes visible in the LDOS of Figure~\ref{fig:Theory}\textcolor{red}{g} are lifted from zero energy, and the emerging corner modes become separated by a gap from the remaining excited states. The predicted energy spectrum for this geometry is displayed in Figure~\ref{fig:CornerMode}\textcolor{red}{a}, while the electronic part of the LDOS of the two corner modes is shown in Figure~\ref{fig:CornerMode}\textcolor{red}{d}, where the enhanced LDOS at the corners is visible.
    
    \subsection{Rhombic lattice}\label{subsect: SI theory B}
\begin{subequations}\label{eq:Normal state Hamilt Rhomb}
    We model the rhombic lattice for the $\mathsf{B}$  terminations with the tight binding superconducting model introduced in Eq.~\eqref{eq:Norm state Hamiltonian B} in the Method section. As for the rectangular lattice, here we consider nearest-neighbor hopping, spin-orbit coupling, and Hund's coupling between electron spin and adatom magnetic moments. In addition, we also include next-to-nearest hopping and spin-orbit coupling terms, to ensure that there are no additional artificial symmetries left in the theoretical model.
    In the momentum-space basis corresponding to $\Psi_{\mathrm{n}\bm{k}}= (\hat{c}_{\bm{k} \uparrow 1}, \hat{c}_{\bm{k} \downarrow 1},\hat{c}_{\bm{k} \uparrow 2}, \hat{c}_{\bm{k} \downarrow 2})$, the normal state Bloch Hamiltonian reads
    \begin{equation}
        H_{\mathrm{n}}(\bm{k}) = \begin{pmatrix}
       J \sigma^z - \mu \sigma^0 + h_{\mathrm{nn}}(k_x, k_y) &   & h_{\mathrm{t}}(k_x, k_y) + h_{\mathrm{soc}}(k_x, k_y)  \\
      h^{\dagger}_{\mathrm{t}}(k_x, k_y) + h^{\dagger}_{\mathrm{soc}}(k_x, k_y) & & -J \sigma^z - \mu \sigma^0 + h_{\mathrm{nn}}(k_x, k_y) 
       \end{pmatrix}
    \end{equation}
    where we defined the hopping and spin orbit coupling matrix elements as follows
    \begin{align}
            h_{\mathrm{t}}(k_x, k_y) &= \sigma^0 t (1+e^{\mathrm{i}k_x}+e^{\mathrm{i}k_y}+e^{\mathrm{i}(k_x+k_y)})\\
             h_{\mathrm{soc}}(k_x, k_y) &=  -\mathrm{i}[(d_x \sigma^y - d_y \sigma^x) -(\sigma^y d_x - d_y\sigma^x) e^{\mathrm{i}k_x}+(\sigma^y d_x + d_y\sigma^x) e^{\mathrm{i}k_y} +(-\sigma^y d_x + d_y\sigma^x) e^{\mathrm{i}(k_x+k_y)} ]\\
             h_{\mathrm{nn}}(k_x, k_y) &=2 (t_x \cos k_x + t_y \cos k_y) \sigma^0 + (-\lambda_x \sin k_x \sigma^y + \lambda_y \sin k_y \sigma^x).
    \end{align}\end{subequations}
    With $s$-wave superconducting pairing, the Bogoliubov de Gennes Hamiltonian in Nambu space becomes
    \begin{equation}\label{eq:BdG H for B - SM}
         H(\bm{k}) = \begin{pmatrix}
        H_{\mathrm{n}}(\bm{k}) & \hat\Delta \\
       \hat{\Delta}^{\dagger} & -H^{*}_{\mathrm{n}}(-\bm{k})
        \end{pmatrix}, \quad \hat{\Delta} = \Delta \begin{pmatrix}
        \mathrm{i}\sigma^y & 0 \\
          0 & \mathrm{i}\sigma^y  
        \end{pmatrix},
    \end{equation}
    where we adopted as basis spinor $\Psi_{\bm{k}}=(\hat{c}_{\bm{k} \uparrow 1}, \hat{c}_{\bm{k} \downarrow 1},\hat{c}_{\bm{k} \uparrow 2}, \hat{c}_{\bm{k} \downarrow 2}, \hat{c}^{\dagger}_{-\bm{k}, 1, \uparrow}, \hat{c}^{\dagger}_{-\bm{k} \downarrow 1},\hat{c}^{\dagger}_{-\bm{k} \uparrow 2}, \hat{c}^{\dagger}_{-\bm{k} \downarrow 2})$. 
    
\begin{figure}[t]
    \centering
    \includegraphics[width=1\textwidth]{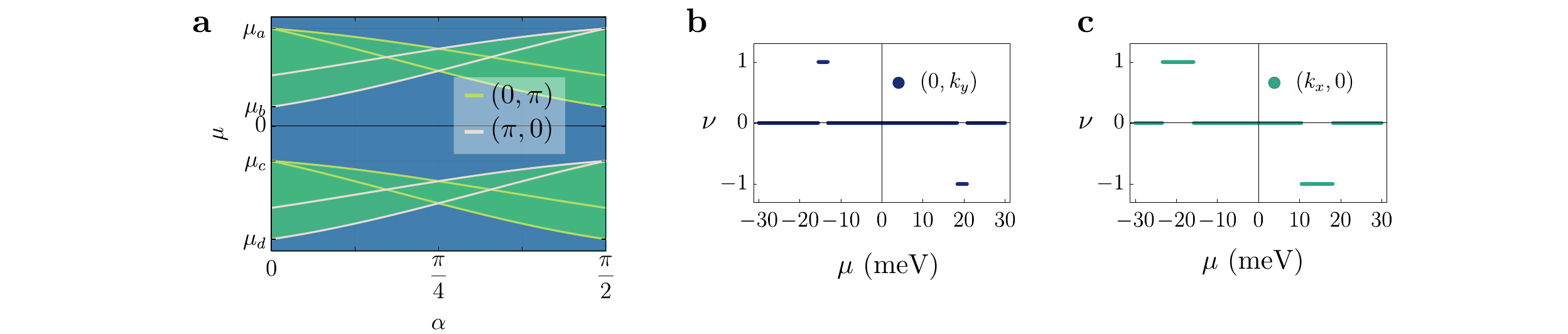}    \caption{\textbf{a} Phase diagram for the rhombic lattice as a function of of chemical potential $\mu$ and anisotropy parameter $\alpha$. Regions shaded in blue correspond to gapped spectrum of the Bogoliubov de Gennes Hamiltonian, while regions colored in green have gapless points along lines with either $k_{x}=\pi$ or $k_{y}=\pi$. The gap closings occur at the high symmetry points $\bm{k}=(0, \pi)$ (black lines) or $\bm{k}=(\pi, 0)$ (yellow lines). Here, $\mu_{a-d}$ only mark the values at the phase boundaries assumed by the critical chemical potential introduced in Eqs.~\eqref{eq:mu crit 1},~\eqref{eq:mu crit 2} at $\alpha=0$. \textbf{b},\textbf{c}: Chiral winding number of the Bogoliubov de Gennes Hamiltonian in case $\mathsf{B}$ for a cut along the phase diagram \textbf{a} with constant $\alpha=0.5$. The paths considered are \textbf{b} $\bm{k}=(0, k)$ and \textbf{c} $\bm{k}=(k, 0)$ for $k\in[-\pi, \pi)$.}
    \label{fig:rhombicSI}
\end{figure}

    In the thermodynamic limit of this model, the bulk is either trivially gapped or has gapless nodal points that are symmetry protected by the crystal symmetries. The gapped phase is topologically trivial, and it occurs for values of the chemical potential $\mu$ far from the value of $J$, the Hund's coupling term which determines the splitting between sets of bands of the normal state Hamiltonian.
    The gap closings occur either at the high symmetry point $\bm{k}=(0, \pi)$, for the value of the chemical potential
    \begin{equation}
        \mu^c_{(0, \pi)} = \pm d_y \pm \sqrt{J^2 - \Delta^2} + 2 (t_x - t_y)
        \label{eq:mu crit 2}
    \end{equation}
    or at the high symmetry momentum $\bm{k}=(\pi, 0)$, for critical chemical potential
    \begin{equation}
        \mu^c_{(\pi, 0)} = \pm d_x \pm \sqrt{J^2 - \Delta^2} - 2 (t_x - t_y).
        \label{eq:mu crit 1}
    \end{equation}
    As for the rectangular lattice, we describe the parameters for the $[001]$ and $[1\Bar{1}0]$ crystallographic axes in terms of the anisotropy parameter $\alpha$. More explicitly
    \begin{equation}
        d_{x} = d \cos\alpha, \ 
        d_{y}=d\sin\alpha, \quad 
        t_{x}=t_{\mathrm{nn}}\cos\alpha,\ t_{y}=t_{\mathrm{nn}}\sin\alpha, \quad 
        \lambda_{x} =\lambda \cos\alpha, \ \lambda_{y} = \lambda\sin\alpha,
        \label{eq:anisotropy rhombic}
    \end{equation}
    where $t_{\mathrm{nn}}$ and $\lambda$ describe the next-to-nearest neighbor hopping and spin-orbit coupling amplitudes respectively.
    The phase diagram as a function of chemical potential $\mu$ and anisotropy $\alpha$, obtained using Eqs.~\eqref{eq:mu crit 1},~\eqref{eq:mu crit 2} and~\eqref{eq:anisotropy rhombic}, is shown in Figure~\ref{fig:rhombicSI}.

    To see why the gapless nodal features are protected by the crystalline symmetries of the lattice, we consider the latter and their action in momentum space.
In the basis of the normal state Hamiltonian, the glide mirrors $M_x$ and $M_y$ are
\begin{equation}
    M_{x, \mathrm{n}}(k_y) = \begin{pmatrix}
    0 & \mathrm{i}\sigma^x e^{i k_y}\\
    \mathrm{i}\sigma^x & 0 
    \end{pmatrix}
    \qquad M_{y, \mathrm{n}} (k_x)= \begin{pmatrix}
    0 & \mathrm{i}\sigma^y e^{i k_x}\\
    \mathrm{i}\sigma^y & 0 
    \end{pmatrix}
\end{equation}
and in Nambu space with spinor $\Psi_{\bm{k}}= (\hat{c}_{\bm{k} \uparrow 1}, \hat{c}_{\bm{k} \downarrow 1},\hat{c}_{\bm{k} \uparrow 2}, \hat{c}_{\bm{k} \downarrow 2}, \hat{c}^{\dagger}_{-\bm{k}, 1, \uparrow}, \hat{c}^{\dagger}_{-\bm{k} \downarrow 1},\hat{c}^{\dagger}_{-\bm{k} \uparrow 2}, \hat{c}^{\dagger}_{-\bm{k} \downarrow 2})$, glide mirrors operations and particle-hole symmetry take the form
\begin{equation}
\begin{split}
       M_{x/y}(k_{y/x}) = \begin{pmatrix}
     M_{x/y, \mathrm{n}}(k_{y/x}) & 0 \\
     0 &  M^*_{x/y, \mathrm{n}}(-k_{y/x}) 
    \end{pmatrix}, \quad
    \mathcal{P}=\begin{pmatrix}
    0 & \mathbb{1} \\
    \mathbb{1} & 0
    \end{pmatrix}
\end{split}
\end{equation}
Starting from these expressions for the glide mirrors, one can see that two eigenstates that are paired by the superconducting coupling are characterized by opposite $M_x$ eigenvalue at $k_x=\pi$, and therefore cannot hybridize and open a gap. Likewise, the eigenstates at $k_y=\pi$ have opposite $M_y$ eigenvalues. 
It is useful to first derive the commutation relations between glide mirror symmetries at momenta $k_x=0, \pi$, which read
\begin{equation}
    \begin{split}
        M_x(0, -k_y)M_y(0, k_y) &= - e^{-\mathrm{i}k_y}M_y(0, k_y)M_x(0, k_y)\\
        M_x(\pi, -k_y)M_y(\pi, k_y) &= e^{-\mathrm{i}k_y}M_y(\pi, k_y)M_x(\pi, k_y)\\
        M_y(-k_x,0)M_x(k_x, 0) &=- e^{-\mathrm{i}k_x}M_x(k_x,0)M_y(k_x,0)\\
        M_y(-k_x,\pi)M_x(k_x, \pi) &= e^{-\mathrm{i}k_x}M_x(k_x,\pi)M_y(k_x,\pi).
    \end{split}
\end{equation}
Let us consider an eigenstate of the glide mirror symmetry $M_{x}$ at $k_x=\pi$, with
\begin{equation}
    M_{x}(\pi, k_y) \ket{\psi^{\pm}(\pi, k_y)} = \pm \mathrm{i} e^{\mathrm{i} \frac{k_y}{2}} \ket{\psi^{\pm}(\pi, k_y)}.
\end{equation}
The state that pairs with $\ket{\psi^{\pm}(\pi, k_y)}$ can be obtained by applying particle-hole symmetry and then $M_y$ to the latter
\begin{equation}\label{eq:state that pairs glide mirror P}
    \ket{\widetilde{\psi}^{\pm}(\pi, k_y)} = M_y(\pi, -k_y) \mathcal{P}\ket{\psi^{\pm}(\pi, k_y)}.
\end{equation}
For the state in~\eqref{eq:state that pairs glide mirror P}, we compute the transformation properties under the action of $M_{x}$, and find
\begin{equation}\label{eq:proof of opposite mirror eval}
    \begin{split}
     M_{x}(\pi, k_y) \ket{\widetilde{\psi}^{\pm}(\pi, k_y)} &=M_{x}(\pi, k_y) M_y(\pi, -k_y) \mathcal{P}\ket{\psi^{\pm}(\pi, k_y)}\\
   &= \mathcal{P}M_{x}(\pi, -k_y) M_y(\pi, k_y) \ket{\psi^{\pm}(\pi, k_y)}\\
   &= \mathcal{P} M_y(\pi, k_y) M_{x}(\pi, k_y) e^{\mathrm{i}k_y}\ket{\psi^{\pm}(\pi, k_y)}\\
   &= \mathcal{P} M_y(\pi, k_y) (\pm \mathrm{i} e^{\mathrm{i}\frac{k_y}{2}}) e^{-\mathrm{i}k_y}\ket{\psi^{\pm}(\pi, k_y)}\\
   &= \mp \mathrm{i} e^{\mathrm{i}\frac{k_y}{2}}  M_y(\pi, k_y)\mathcal{P} \ket{\psi^{\pm}(\pi, k_y)} =  \mp \mathrm{i} e^{\mathrm{i}\frac{k_y}{2}} \ket{\widetilde{\psi}^{\pm}(\pi, k_y)}.
   \end{split}
\end{equation}
Eq.~\eqref{eq:proof of opposite mirror eval} implies that the two eigenstates connected by the superconducting pairing remain degenerate, as they carry opposite $M_x$ eigenvalue.
Following the same procedure, one can show that the two eigenstates of $M_x$ that pair at $k_x=0$ have the same mirror eigenvalue, and therefore the degeneracy between the two eigenstates can be lifted, in accordance with the absence of gapless points along the high symmetry line $(0, k_y)$. With the same arguments, one can see that eigenstates paired by superconductivity at $k_y=\pi$ carry opposite $M_y$ eigenvalue, while they have the same eigenvalue at $k_y=0$.

The composition of time reversal operation
\begin{equation}
    \mathcal{T} = \begin{pmatrix} 0 & \mathrm{i}\sigma^y \\
    \mathrm{i}\sigma^y & 0
    \end{pmatrix} \mathcal{K},
\end{equation}
where $\mathcal{K}$ indicates complex conjugation,
and standard mirror symmetries $M'_{x/y}$ leads to an effective time reversal symmetry $\widetilde{M}$ with $\widetilde{M}^2 = (\mathcal{T}M_{x/y}')^2= +\mathbb{1}$. Therefore, chiral symmetry $\mathcal{C}=\widetilde{M}\mathcal{P}$ is well defined and we can compute the chiral topological invariant of class BDI for any gapless particle-hole invariant path, see Figure~\ref{fig:rhombicSI}\textcolor{red}{b},\textcolor{red}{c}.

In the numerical plots appearing throughout the Supplementary Information and the main text, we used the parameters $J=17$~meV, $\Delta=1.5$~meV, $t=4.0$~meV, $\mu=20$~meV, $\alpha=0.5$, $d=1$~meV, $t_{\mathrm{nn}}=2$~meV, $\lambda=0.1$~meV and thermal broadening $\varepsilon=0.1$~meV, used in the calculation of the LDOS maps at zero energy.

\subsection{Importance of spin-orbit coupling}
\begin{figure}[t]
    \centering
    \includegraphics[width=1\textwidth]{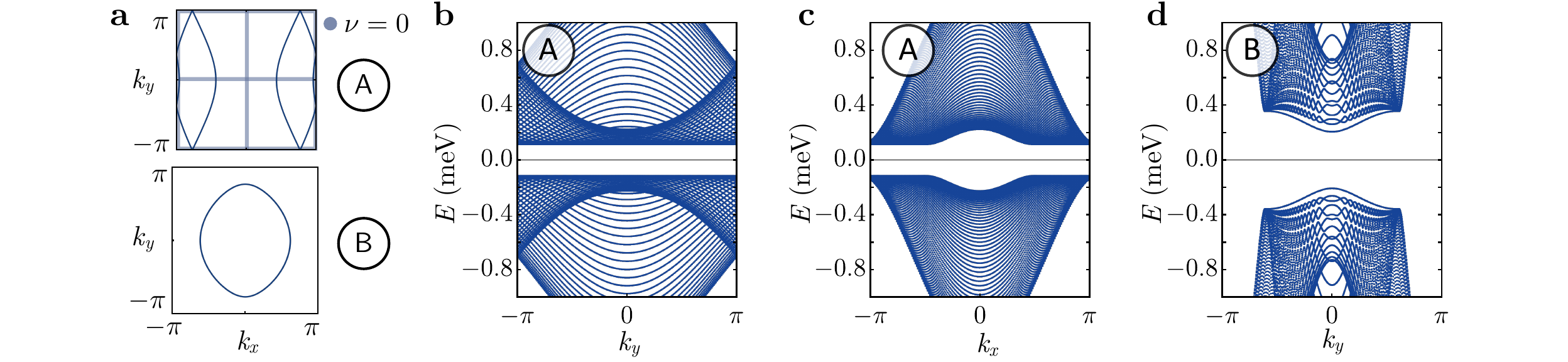}
    \caption{\textbf{Absence of spin-orbit coupling:} \textbf{a} Normal state Fermi surface for the rectangular (upper figure) and rhombic (lower figure) lattices. The high symmetry lines in the BZ of the rectangular lattice are colored according to the value of the topological invariant, which is everywhere trivial. \textbf{b}-\textbf{d}: Ribbon spectra for the rectangular lattice with open boundary conditions along \textbf{b} $x$ and \textbf{c} $y$, and ribbon spectra for the rhombic lattice \textbf{d} with open boundary conditions along $x$.}
    \label{fig:noSOC_SM}
\end{figure}

As an additional comment on the discussions presented in Sec.~\ref{subsect: SI theory A} and Sec.~\ref{subsect: SI theory B}, we shortly consider the case of no spin-orbit coupling in the models for the Shiba lattices.
When spin-orbit coupling is absent, the the two type of lattices, both rectangular and rhombic, are gapped and have no non-trivial topological features.
Figure~\ref{fig:noSOC_SM} shows the Fermi surface and the ribbon spectra corresponding to the ones shown in Figure~\ref{fig:Theory}, with all the spin-orbit coupling strength parameters set to zero.

The prediction of a gapless bulk for the terminations $\mathsf{A}$ and $\mathsf{B}$, in the absence of spin-orbit coupling, is in contrast with the experimental observations. In fact, Figure~\ref{fig:Spectroscopy} hints at a non-zero LDOS close to zero bias in all the measured terminations, either for the the edges of the structures or in their bulk.

\subsection{Discussion of adatom spins with in-plane component}\label{sec:Non z-aligned spins}

\begin{figure}[t]
    \centering
    \includegraphics[width=1\textwidth]{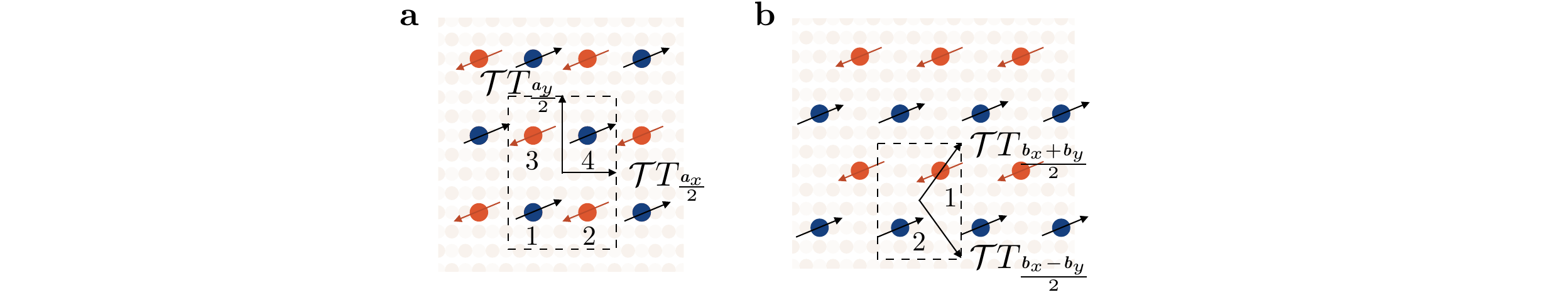}
    \caption{Antiferromagnetic ordering of the spins in \textbf{a} the rectangular and \textbf{b} rhombic lattice. The choice of unit cell is highlighted by the dashed rectangles, and the half-lattice traslations discussed in the text are marked by black arrows. Here colored dots indicate the $z$ component of the spins either up (red) or down (blue), and the arrows indicate the in-plane projection of the spins. Red (blue) sites correspond to spins with $+\bm{S}$ ($-\bm{S}$) in Eq.~\eqref{eq:non z spins direction}.}
    \label{fig:nonzSpins}
\end{figure}

In this section, we discuss the consequences of relaxing the assumption of adatom spins aligned along the $z$-direction. Instead, we admit an in-plane component of the spin. Based on the experimental measurements on 1D chains and on the DFT calculations discussed in the main text, we assume that the antiferromagnetic ordering is maintained. For this, we parametrize the adatom spins by
\begin{equation}\label{eq:non z spins direction}
    \pm \bm{S} = \pm S(\bm{e}_z \cos\theta + \bm{e}_x \sin\theta\cos\phi + \bm{e}_y \sin\theta\sin\phi)
\end{equation}
with the usual definition of spherical angles $\phi \in[0, 2\pi)$ and $\theta \in [0, \pi)$, and where the alternating sign should reproduce the antiferromagnetic pattern.
Figure~\ref{fig:nonzSpins} shows the scheme for the magnetic ordering of the two lattices considered.

In practice, this amounts to replacing the terms $\pm J\sigma^z$ in the Hamiltonians~\eqref{eq:Normal state Hamilt Rect} and~\eqref{eq:Normal state Hamilt Rhomb} by
\begin{equation}\label{eq:non-z spins}
   \pm J \sigma^z \rightarrow \pm J (\sigma^z \cos\theta + \sigma^x \sin\theta\cos\phi + \sigma^y \sin\theta\sin\phi).
\end{equation}
In the following, we denote Hamiltonians with the replacement~\eqref{eq:non-z spins} by $H_{S, \text{n}}(\bm{k})$ and $H_{S}(\bm{k})$ for the normal state and Bogoliubov de Gennes Hamiltonians respectively. 
\vspace{10pt}
\paragraph{Rectangular lattice}
First, we focus on the case of the rectangular lattice, relevant for the termination $\mathsf{A}$.
For a small tilting of the spins, i.e., $\theta \ll 1$, the phase diagram shown in Figure~\ref{fig:PhaseDiagramA}\textcolor{red}{a} remains qualitatively unchanged, apart from a renormalization of the boundaries between different phases.
With the replacement~\eqref{eq:non-z spins}, the mirror operations $M_{x/y}$ and the glide mirror operations combined with time reversal symmetry $\widetilde{M}_{x/y}$ are no longer symmetries of the lattice.
Nevertheless, there are two crystalline symmetries left, which are defined by the composition of time reversal symmetry and half a lattice translation in either the $x$ or $y$ direction, see Figure~\ref{fig:nonzSpins}\textcolor{red}{a}.
These two operations, denoted by $\widetilde{T}_{x/y} = \mathcal{T} T_{\bm{a}_{x/y}/2}$, define an effective time reversal symmetry in the crystal.
In this case $\widetilde{T}_{x}^2 = -e^{\mathrm{i}k_x} \mathbb{1}$ and $\widetilde{T}_{y}^2 = -e^{\mathrm{i}k_y} \mathbb{1}$, and the unitary part of these operations acts on the normal state Hamiltonian as
\begin{equation}
    \widetilde{T}_{x, \text{n}}(k_x)= \begin{pmatrix}
    0 & \mathrm{i}\sigma^y & 0 & 0 \\
    \mathrm{i} \sigma^y e^{\mathrm{i} k_x} & 0 & 0 & 0 \\
    0 & 0 & 0 & \mathrm{i}\sigma^y\\
    0 & 0 & \mathrm{i}\sigma^y e^{\mathrm{i}k_x} & 0\\
    \end{pmatrix}, \qquad 
    \widetilde{T}_{y,\text{n}}(k_y) = \begin{pmatrix}
    0 & 0 & \mathrm{i}\sigma^y e^{-\mathrm{i} k_y} & 0\\
    0 & 0 & 0 & \mathrm{i} \sigma^y e^{-\mathrm{i} k_y}\\
    \mathrm{i}\sigma^y & 0 & 0 & 0 \\
    0 & \mathrm{i}\sigma^y & 0 & 0\\
    \end{pmatrix}
\end{equation}
The extension to Nambu space of the latter reads
\begin{equation}
    \widetilde{T}_{x/y}(k_{x/y}) = \begin{pmatrix}
    \widetilde{T}_{x/y, \text{n}}(k_{x/y}) & 0 \\
    0 & \widetilde{T}^*_{x/y, \text{n}}(-k_{x/y})
    \end{pmatrix},
\end{equation}
and the action on the Bogoliubov de Gennes Hamiltonian is
\begin{equation}
    \widetilde{T}_{x}(k_{x}) H^*_S(k_x, k_y) \widetilde{T}^{-1}_{x}(k_{x}) = H_S(-k_x, -k_y), \qquad \widetilde{T}_{y}(k_{y}) H^*_S(k_x, k_y) \widetilde{T}^{-1}_{y}(k_{y}) = H_S(-k_x, -k_y).
\end{equation}
Particle-hole symmetry is still a valid symmetry of the system, hence we can still assign the high symmetry lines in the BZ to some of the classes in the tenfold way.
More precisely, the lines parametrized by $\bm{k}=(0, k)$ and $\bm{k}=(k, 0)$, with $k \in[0, 2\pi)$, belong to the DIII class of the tenfold way, while $\bm{k}=(\pi, k)$ and $\bm{k}=(k, \pi)$, for $k \in[0, 2\pi)$, belong to the BDI class.
This allows to define the topological invariants $\tilde{\nu}_{x, k_x=0}, \tilde{\nu}_{y, k_y=0}\in \mathbb{Z}_2$ for the lines in class DIII and  $\nu_{x, k_x=\pi}, \nu_{y, k_y=\pi} \in \mathbb{Z}$ for the lines in class BDI. The topological invariants $\nu$ and $\tilde{\nu}$ can be computed from the chiral winding number~\eqref{eq:chiral winding number} and the Wilson loop~\eqref{eq:Wilson Loop} respectively.
From the evaluation of these invariants for a range of parameters, we obtain that $\nu_{x, k_x}$ and  $\nu_{y, k_y}$ always remain trivial, irrespective of the value of the chemical potential $\mu$, while $\tilde{\nu}_{x, k_x} = 1$ and  $\tilde{\nu}_{y, k_y}=0$ in the topological phases corresponding to $2_x$ and $2_y$. Hence, even in the case of a generic adatom spin direction, there are two topological zero modes protected by the effective time-reversal symmetry and particle-hole symmetry, that appear along the $y$-edges of an open boundary termination analogous to $\mathsf{A}$. These topological modes are still protected by a lattice symmetry, in particular the half-lattice translation $T_{\bm{a}_y/2}$, which is still respected at the $y$-boundaries of $\mathsf{A}$. Therefore, this system realizes a topological crystalline insulator as long as antiferromagnetic ordering is ensured, irrespective of the direction of the adatom spins.

Although the topological boundary modes are stable when choosing~\eqref{eq:non z spins direction}, the arguments for the higher order topological modes described in Sec.~\ref{subsec:HigherOrderTSC} fail in the case of a generic adatom spin orientation. In fact, the condition for the realization of the corner mode at zero energy relies on the corner being invariant under the action of the mirror operation $\widetilde{M}_{y}$, which is no longer a symmetry of $H_S(\bm{k})$.
\vspace{10pt}
\paragraph{Rhombic lattice}
For the case of the rhombic lattice, once more the mirror operations $M_{x/y}$ and $\widetilde{M}_{x/y}$ are no longer symmetries of the lattice once a generic adatom spin orientation is considered.
Under the assumption of antiferromagnetic ordering, the system retains the invariance under the combination of a half-lattice translation in each direction, $T_{(\bm{b}_x \pm \bm{b}_y)/2}$ combined with time-reversal symmetry, see Figure~\ref{fig:nonzSpins}\textcolor{red}{b}. In the following, we only consider $T_{(\bm{b}_x + \bm{b}_y)/2}$ as the two translations are equivalent up to a global phase factor. This half-lattice translation defines an effective time-reversal symmetry $\widetilde{T}= \mathcal{T}T_{(\bm{b}_x + \bm{b}_y)/2}$.
This operation can be written as
\begin{equation}
   \widetilde{T}_{\text{n}}(\bm{k}) = \begin{pmatrix}
   0 & \sigma^y e^{-\mathrm{i} \frac{(k_x+ k_y)}{2}} \\
  \sigma^y e^{\mathrm{i} \frac{(k_x+k_y)}{2}} & 0
   \end{pmatrix}, \qquad \widetilde{T}(\bm{k}) = \begin{pmatrix}
   \widetilde{T}_{\text{n}}(\bm{k}) & 0 \\
   0 & -\widetilde{T}^*_{\text{n}}(-\bm{k})
   \end{pmatrix}
\end{equation}
and the action of the Bogoliubov de Gennes Hamiltonian is
\begin{equation}
    \widetilde{T}(\bm{k}) H^*_S(\bm{k})  \widetilde{T}^{-1}(\bm{k}) =  H_S(-\bm{k}) .
\end{equation}
The combination of $\widetilde{T}$ with particle-hole symmetry $\mathcal{P}$ results in a unitary operator $U(\bm{k})= \widetilde{T}(-\bm{k})\mathcal{P}$ that anticommutes with the Hamiltonian
\begin{equation}\label{eq:anticomm U}
    U(\bm{k}) = \begin{pmatrix}
    0 & 0 & 0 & \sigma^y e^{\mathrm{i}\frac{(k_x+ k_y)}{2}} \\
    0 & 0 & \sigma^y e^{-\mathrm{i}\frac{(k_x+ k_y)}{2}} & 0 \\
    0 & \sigma^y e^{\mathrm{i}\frac{(k_x+ k_y)}{2}} & 0 & 0 \\
    \sigma^y e^{-\mathrm{i}\frac{(k_x+ k_y)}{2}} & 0 & 0 & 0 \\
    \end{pmatrix}, \quad U(\bm{k}) H(\bm{k}) U^{-1}(\bm{k}) = - H(\bm{k}),
\end{equation}
hence realizing an effective chiral symmetry.
This anticommutation relation protects the nodal points, although it does not constrain them to be fixed at the high symmetry lines with $k_x=\pi$ or $k_y=\pi$, as it was the case in Sec.~\ref{subsect: SI theory B}. 
In fact, adding a term to the Hamiltonian
\begin{equation}
    H_S(\bm{k}) \rightarrow  H_S(\bm{k}) + U(\bm{k})
\end{equation}
would open a gap in the bulk energy spectrum, but would also violate the condition~\eqref{eq:anticomm U}.
Hence, for a small tilting of the adatom spin direction, meaning $\theta \ll 1$ in~\eqref{eq:non z spins direction}, the bulk is still gapless and has nodal points close to the high symmetry lines at $k_x=\pi$ or $k_y=\pi$, where they are pinned for adatom spins aligned along the out-of-plane direction.

\subsection{Effects of a magnetic field on symmetries}
Here we shortly discuss the effects that the introduction of an external magnetic field has on the symmetries of the tight-binding models discussed above. We assume that the magnetic field remains small enough, such that the adatom spins are still antiferromagnetically ordered and pointing along the out-of-plane direction and that the superconductivity in the Nb substrate remains substantially unperturbed (i.\,e. without a significant reduction of the size of the superconducting gap). As a first level approximation, we neglect Zeeman and orbital effect, and only discuss the changes occurring in the list of symmetries of the systems.

Let us first consider the rectangular lattice $\mathsf{A}$.
The introduction of a magnetic field pointing along the out-of plane direction $z$, corresponding to the $[110]$ crystallographic axis of Nb, does not break any of the relevant symmetries of the model in Eqs.~\eqref{eq:Normal state Hamilt Rect},~\eqref{eq:BdG H for A - SM}.
In fact, the magnetic field remains invariant under the combination of one of the mirror operations $M_{x/y}$ and time reversal, $\widetilde{M}_{x/y}$.
Hence, the analysis remains substantially unchanged for this scenario, until the magnetic field becomes strong enough to modify the antiferromagnetic adatom spin pattern.
Instead, the presence of a magnetic field directed along the $x$ axis ($[001]$) breaks the symmetry $\widetilde{M}_{x}$, while it leaves intact the symmetry $\widetilde{M}_{y}$.
On the other hand, a magnetic field pointing along the $y$ direction ($[1\bar{1}0]$) preserves the $\widetilde{M}_{x}$ symmetry and breaks the $\widetilde{M}_{y}$ symmetry. Therefore, in the topological phase of the lattice $\mathsf{A}$ we expect that having a magnetic field pointing along either the $z$ or $x$ direction will leave the topological modes unchanged, while a magnetic field along $y$ would perturb the gapless edge modes appearing on the boundaries of the $\mathsf{A}_1$ termination.

The rhombic lattice $\mathsf{B}$ is theoretically described by the model proposed in Eqs.~\eqref{eq:BdG H for B - SM},~\eqref{eq:Normal state Hamilt Rhomb}. For this case, the presence of a magnetic field affects the properties of the model irrespective of the direction of the magnetic field.
An in-plane magnetic field along $x$ would break the glide mirror symmetry $M_x$ and the spatio-temporal $\widetilde{M}_{y}$ while preserving $M_y$ and $\widetilde{M}_{x}$, and vice versa for a magnetic filed pointing along $y$. An out-of-plane magnetic field would violate both the glide mirror symmetries at the same time.
As both the glide mirror symmetries are necessary to protect the gapless points in the bulk spectrum of $\mathsf{B}$, any in-plane or out-of-plane magnetic field will lift the degeneracy of the gapless points and lead to a gap opening, or just allow this points to move to lower-symmetry lines in the BZ while still remaining gapless.
For the spatio-temporal symmetries $\widetilde{M}_{x/y}$, the analysis is identical to the case of the rectangular lattice.

\subsection{Modelling higher energy Shiba orbitals}
Let us now consider tight-binding models analogous to the ones presented in the main text, but where we replace the $d_{z^2}$ orbitals by orbitals with $d_{xy}$ character.
The $d_{z^2}$ orbitals transform under the $A_1$ representation of the site symmetry group of the site at which the Cr adatoms lie, which is isomorphic to the point group $C_{2v}$. This implies that the $d_{z^2}$ orbitals transform trivially under the symmetries that leave the site invariant. On the other hand, the $d_{xy}$ orbitals transform under the $A_2$ representation, hence they transform evenly under the two-fold rotation $C_2$, but oddly under the mirror operations with mirror planes coinciding with either the $x$-$z$ or $y$-$z$ plane. 

Tight-binding models based on the $d_{xy}$ orbitals therefore remain substantially similar to those proposed for $d_{z^2}$ orbitals, except for the change in the sign of some of the tunneling terms, which can now assume negative values.
In the case of the $\mathsf{A}$ lattice, one has to replace all the hopping and spin-orbit-coupling amplitudes in the model devised for the  by $d_{z^2}$ by negative values, as the overlap between neighboring orbitals results into a tunneling amplitude with a negative sign. 
The overlap between next-to-nearest neighbor orbitals would once more be positive, but this higher order correction term is not included in the model for $\mathsf{A}$.
This change in the amplitude sign leads to a substantially similar phase diagram as compared to the one discussed in App.~\ref{subsect: SI theory A}, with the exception that the winding number $\nu$ for the BDI class changes sign.
Hence, considering a model with $d_{xy}$ orbitals alone does not lead to novel predictions in terms of topological modes in the terminations. On the other hand, a two orbital model would lead to a richer phase diagram, with the downside of having a much larger set of parameters. This in principle would allow to realize a range of topological phases with BDI invariants varying from $\nu=-4, \cdots, 4$, which would include phases matching with our current single-band model predictions, as well as more complex topological phases.

\clearpage
\newpage

\include{ExtendedData}

\end{document}

%% file: ExtendedData.tex
% \documentclass[aps,prb,showpacs,preprintnumbers,singlecolumn,superscriptaddress]{revtex4-2}
% \usepackage{amsmath,amssymb}
% \usepackage{bm}
% \usepackage{tipa}
% \usepackage{upgreek}
% \usepackage{comment}
% \usepackage{mathrsfs}
% \usepackage{graphicx}
% %\usepackage{subfig}
% \usepackage{braket}
% \usepackage{enumitem}
% \usepackage{mathbbol}
% \usepackage{booktabs}
% %\usepackage{multibib}
% \usepackage{gensymb}
% \usepackage[normalem]{ulem}
% \usepackage{color}
% \usepackage[colorlinks,bookmarks=true,citecolor=blue,linkcolor=red,urlcolor=blue]{hyperref}
% \usepackage{hyperref}
% \renewcommand{\vec}[1]{\mathbf{#1}}
% \usepackage{pifont}
% \newcommand{\cmark}{\ding{51}}
% \newcommand{\xmark}{\ding{55}}

% \allowdisplaybreaks

% \usepackage{siunitx}
% \usepackage{soul}
\renewcommand{\thefigure}{E\arabic{figure}}

%\begin{document}
\begin{center}
		\textbf{\large --- Extended Data ---\\}
		\medskip
    Martina O.\ Soldini$^1$, Felix K{\"u}ster$^2$, Glenn Wagner$^1$, Souvik Das$^2$, Amal Aldarawsheh$^{3,4}$, 
    
    Ronny Thomale$^{5, 6}$, Samir Lounis$^{3, 4}$, Stuart S.\ P.\ Parkin$^2$, Paolo Sessi$^2$, Titus Neupert$^1$.
	
	\medskip
	\textit{
	$^1$University of Zurich, Winterthurerstrasse 190, 8057 Zurich, Switzerland,\\
	$^2$Max Planck Institute of Microstructure Physics, Halle, Germany,\\
	$^3$Peter Grünberg Institut and Institute for Advanced Simulation, Forschungszentrum Jülich $\&$ JARA, Jülich, Germany.\\
	$^4$Faculty of Physics, University of Duisburg-Essen and CENIDE, Duisburg, Germany.\\
     $^5$ Institut fur Theoretische Physik und Astrophysik, Universitat Wurzburg, Wurzburg, Germany.\\
	$^6$Department of Physics and Quantum Centers in Diamond and Emerging Materials (QuCenDiEM) Group, Indian Institute of Technology Madras, Chennai, India.
	}

	\end{center}
    \subsection{Figure 1}
      \begin{figure}[!h]
        \centering
        \includegraphics[width=0.55\textwidth]{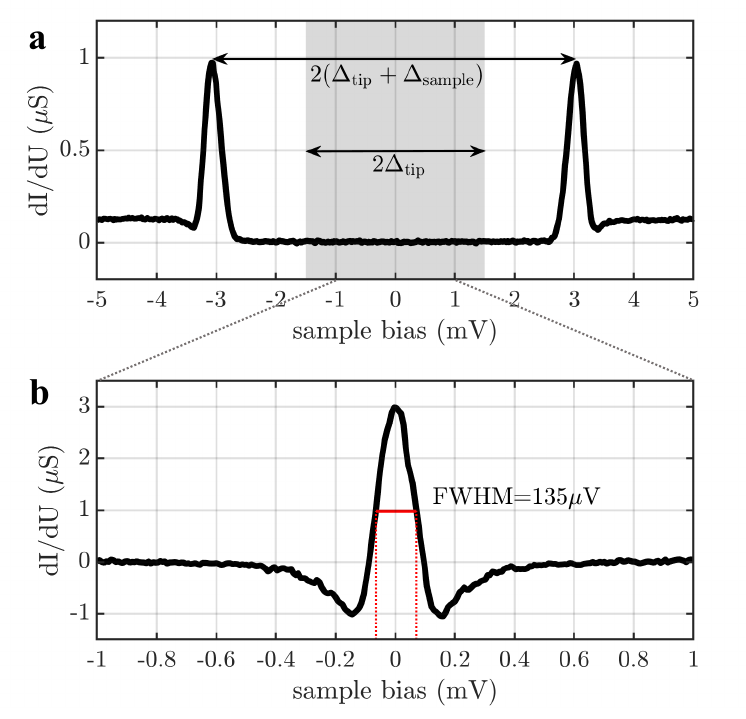}
        \caption{\textbf{Scanning tunneling spectroscopy measured with a superconducting Nb tip on clean Nb(110).} A bulk superconducting Nb cluster at the tip was obtained by deep indentation into the sample. \textbf{a} Tunneling spectroscopy shows the convolution of tip and sample density of states resulting in a large gap with size $2(\Delta_\mathrm{tip}+\Delta_\mathrm{sample})/e$. From the measured convoluted gap size of \SI{3.05}{\milli\volt} and the database value for bulk Nb ($\Delta_\mathrm{Nb}=1.52$ meV), we obtain a tip gap approximately equal to the bulk value, i.e. $\Delta_\mathrm{tip}\approxeq\Delta_\mathrm{Nb}$, which is indicated by the gray area. Measurement parameters: stabilized at sample bias \SI{-5}{\milli\volt} \textbf{b}, tunneling current \SI{500}{\pico\ampere}, bias AC modulation amplitude \SI{40}{\micro\volt}, temperature \SI{500}{\milli\kelvin}. 
        \textbf{b} Zero bias differential conductance peak. The measured FWHM of \SI{135}{\micro\volt} provides a reference for our experimental energy resolution. Measurement parameters: stabilized at sample bias \SI{-5}{\milli\volt}, tunneling current \SI{30}{\nano\ampere}, bias AC modulation amplitude \SI{10}{\micro\volt}, temperature \SI{500}{\milli\kelvin}.}
        \label{fig:EnergyResolution_SM}
    \end{figure}

\newpage
\subsection{Figure 2}
    \begin{figure}[!h]
        \centering
        \includegraphics[width=0.4\textwidth]{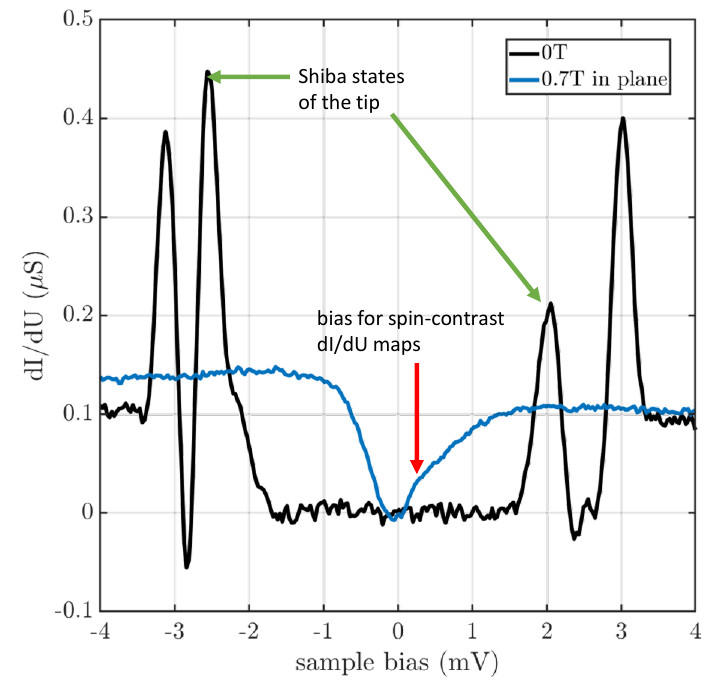}
        \caption{\textbf{Spectroscopy with spin-sensitive tip used for Figure 1e,g of the main text.} Both measurements are acquired on the clean Nb substrate with a superconducting tip featuring single Cr atoms at the apex that induce Shiba states indicated by green arrows for the zero-field measurement (black curve). The applied external magnetic field of \SI{0.7}{\tesla} in-plane overcomes the second critical field for the Nb sample while superconductivity is maintained in the Nb cluster at the tip (blue curve) where Shiba states are still visible as shoulders on the flank of the gap. The observed spin contrast in constant-height d$I$/d$U$ maps was strongest for the indicated bias at \SI{0.33}{\milli\volt}.}
        \label{fig:spinfunctip}
    \end{figure}    
\newpage
\subsection{Figure 3}
        \begin{figure}[!h]
        \centering
        \includegraphics[width=0.60\textwidth]{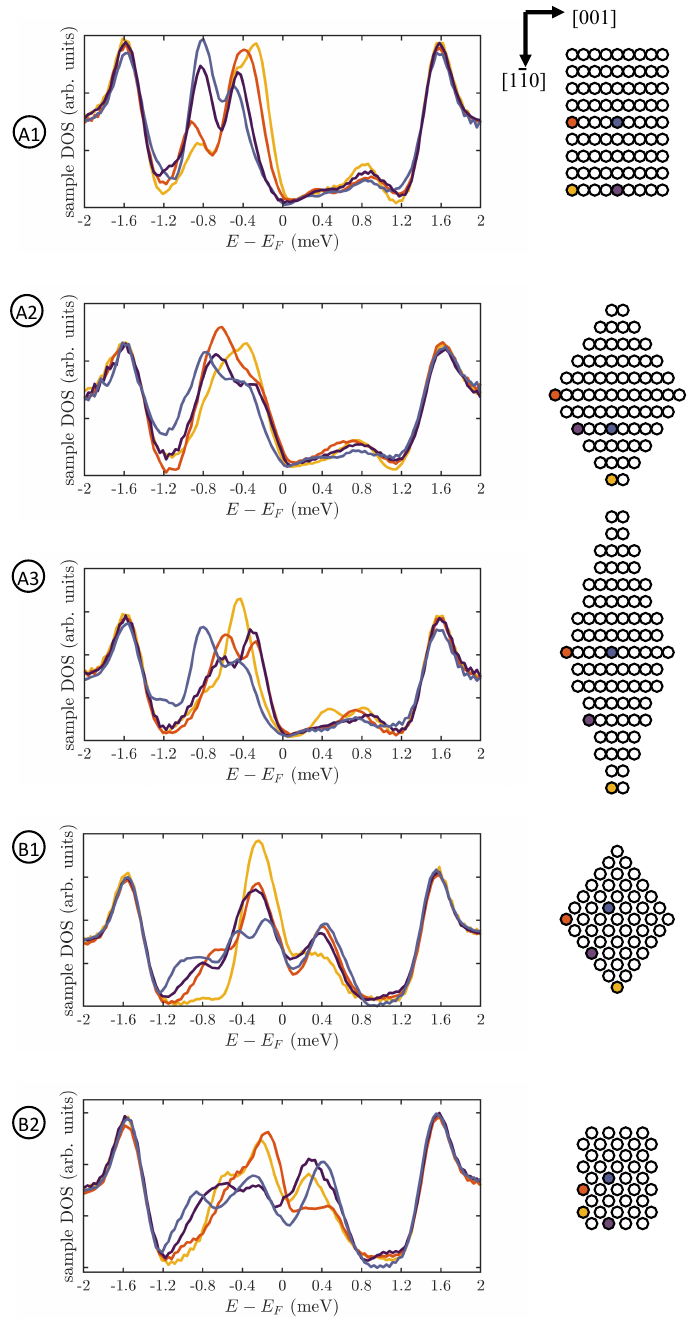}
        \caption{Tunneling spectroscopy data corresponding to figure 2 of the main article after performing numerical deconvolution. Respective lattice positions are indicated by colored circles corresponding to STS lines. Numerical deconvolution was carried out following the process described in~\cite{Schneider2021}.}
        \label{fig:Deconvoluted_SM}
\end{figure}

\newpage
    \subsection{Figure 4}
      \begin{figure}[!h]
        \centering
        \includegraphics[width=0.7\textwidth]{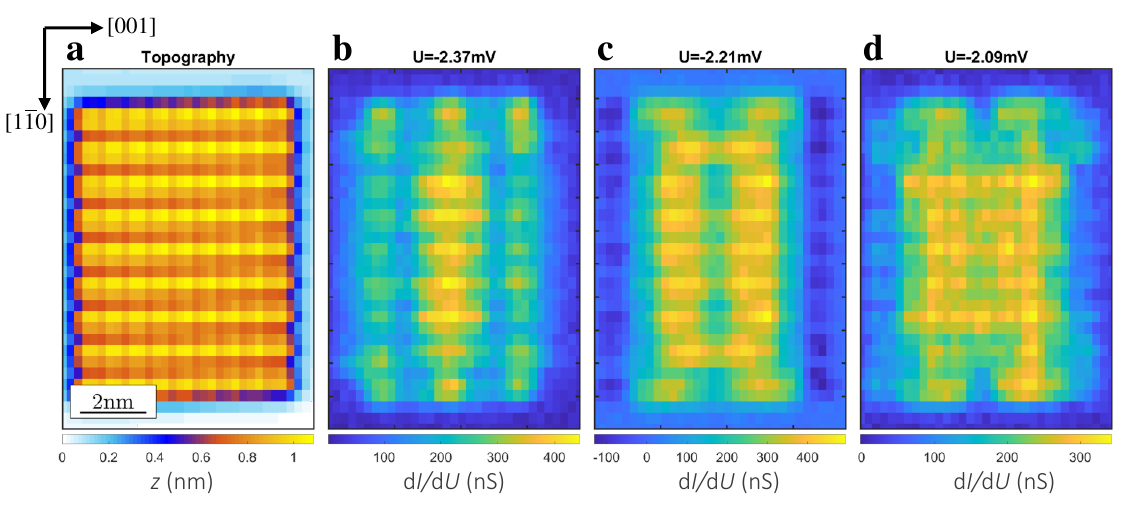}
        \caption{\textbf{Formation of two dimensional Shiba bands inside the bulk of artificial spin structures: a} Topographic image from the full spectroscopic measurement grid on structure \textsf{A$_1$}. \textbf{b,c} Two representative d$I$/d$U$ maps for bulk Shiba bands at sample biases \SI{-2.37}{\milli\volt}, \SI{-2.21}{\milli\volt} and \SI{-2.09}{\milli\volt}, demonstrating their 2D character by intensity modulations in both orthogonal directions ($[001]$ and $[1\bar{1}0]$).}
        \label{fig:BulkEnergyDependence_SM}
    \end{figure}

\newpage
\subsection{Figure 5}
  \begin{figure}[h]
    \centering
    \includegraphics[width=0.75\textwidth]{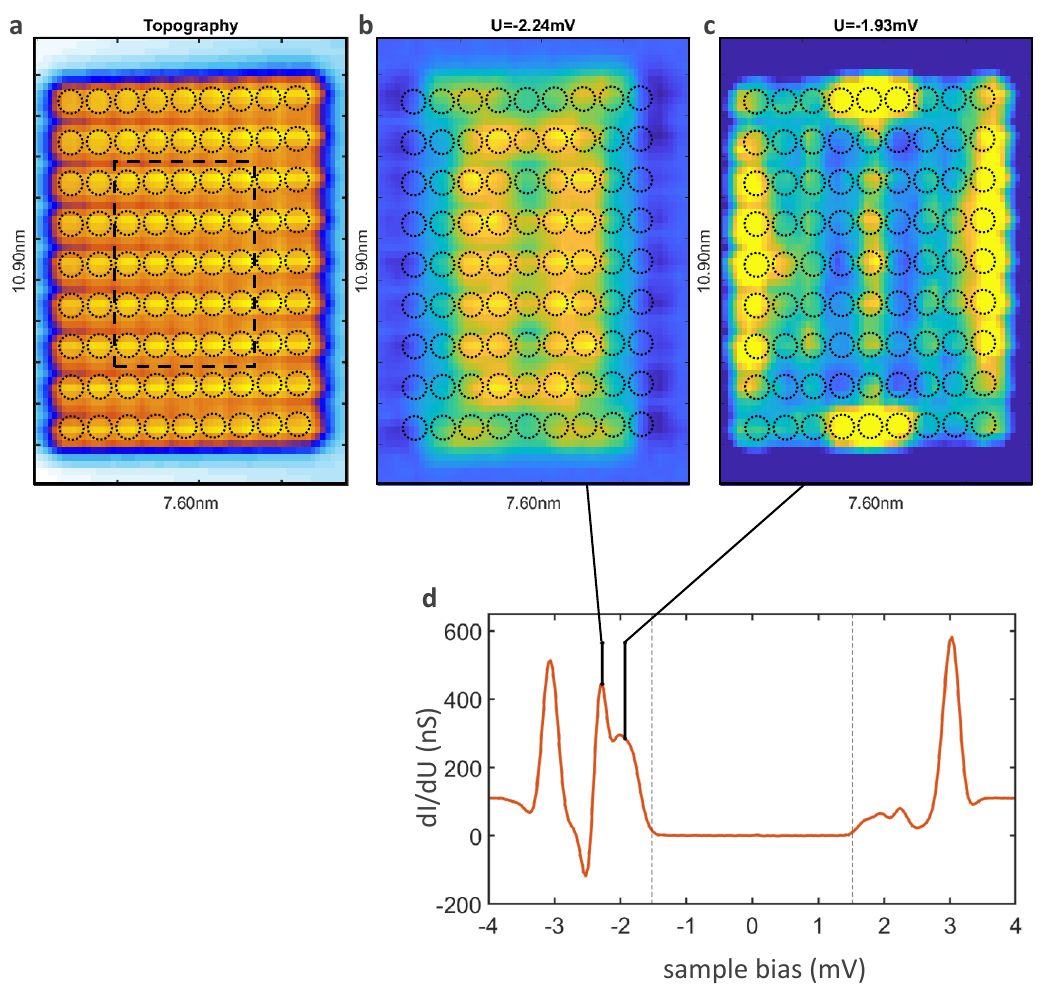}
    \caption{\textbf{Analysis of Shiba orbital characters.}\textbf{a} Topographic image of the $\mathsf{A1}$ lattice analyzed in the main text. \textbf{b-c} d$I$/d$U$ maps acquired at  biases corresponding to the two most prominent peaks in the d$I$/d$U$ spectrum reported in \textbf{d}. The spectrum has been obtained by averaging over the region identified by a dashed black box in \textbf{a}. Two peaks are dominating the scene inside the superconducting gap, centered at $U = -2.24 mV$ and $U = -2.24 mV$ (see black lines). Their strong intensity allow assigning them to $d^2_z$-derived states which can be effectively measured by STS  because of their large extension into the vacuum. This orbital assignment is corroborated by the respective  d$I$/d$U$ maps, which show two distinct patterns, both characterized by intensity maxima centered around the position of the adatoms, as expected for $d^2_z$-derived states. }
    \label{fig:OrbitalAssignment}
\end{figure}

\newpage
\subsection{Figure 6}
        \begin{figure}[!h]
        \centering
        \includegraphics[width=0.6\textwidth]{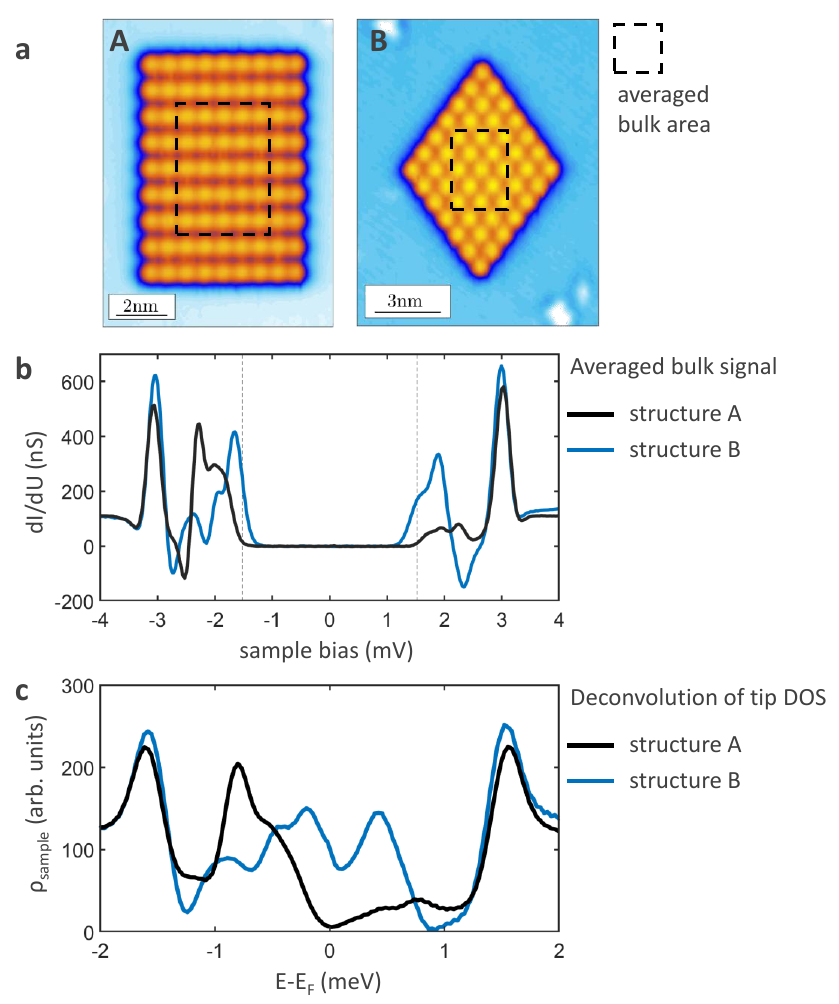}
        \caption{\textbf{Averaged bulk signals of structures A and B. a} Topography images of structures \textsf{A} and \textsf{B} with the respective bulk area indicated by a dashed rectangle used for creating a bulk STS signal which is shown in \textbf{b}. \textbf{c} plots each signal after deconvolution by a model tip DOS corresponding to a Nb superconducting gap in order to get a good approximation of the sample DOS $\rho_\mathrm{sample}$. Structure \textsf{B} has clearly no gap in the bulk while \textsf{A} features a dip at zero energy which is very close to zero intensity. 
        %\textbf{d} Fit of the deconvolution of bulk signal A with four Gaussian peaks (orange). To account for the broadening by finite experimental energy resolution the FWHM of the peaks used for fitting was reduced by \SI{140}{\micro\volt} (green). This results in a clear gap in the bulk of Structure A.
        }
        \label{fig:averagedbulk}
    \end{figure}

\newpage

\subsection{Figure 7}
        \begin{figure}[!h]
        \centering
        \includegraphics[width=0.45\textwidth]{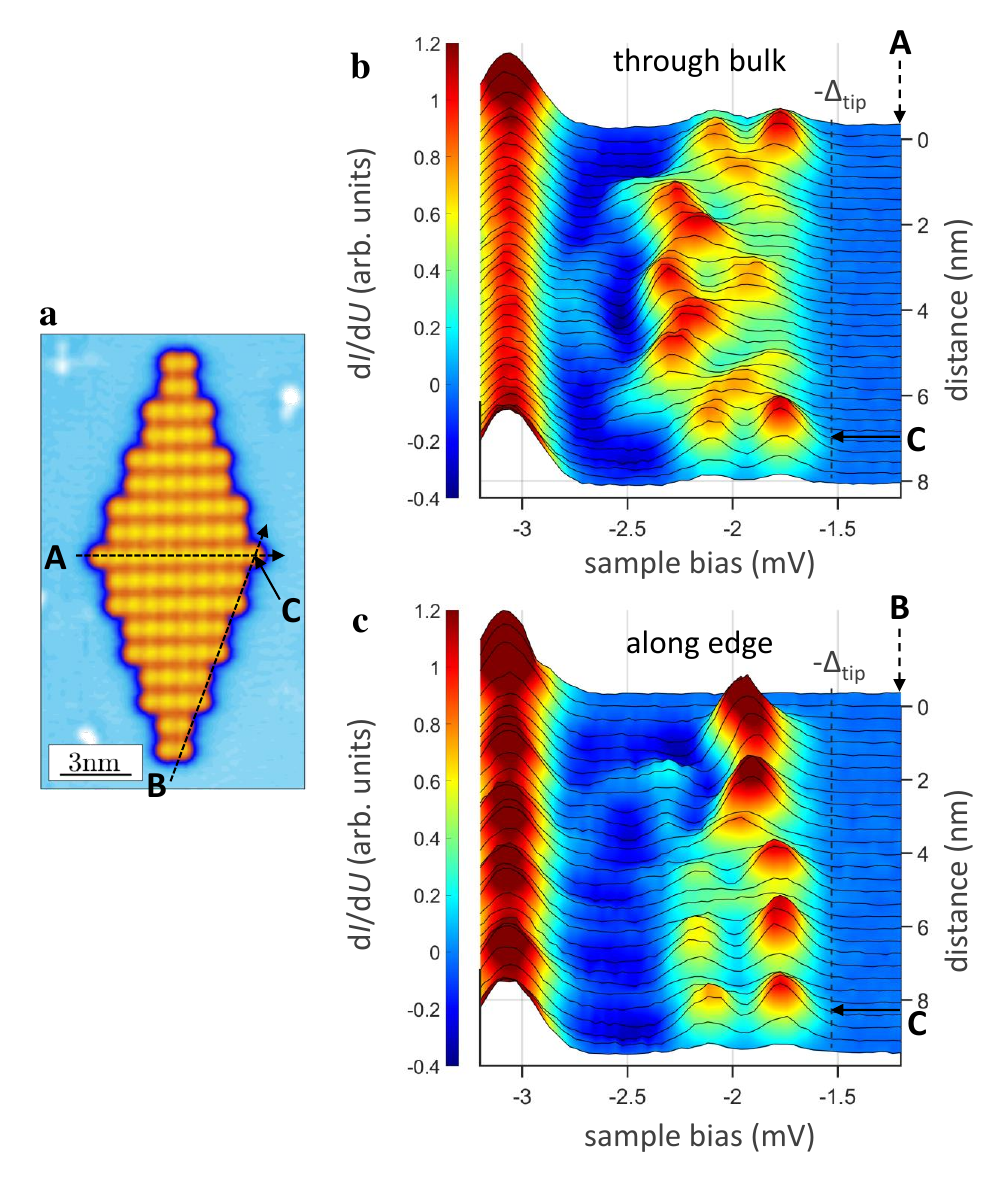}
        \caption{\textbf{Tunneling spectroscopy data acquired in several points along straight lines over the \textsf{A$_3$} termination. a} Topography image of structure \textsf{A$_3$} with two dashed lines indicating STS line paths \textsf{\textbf{A}} (through bulk) and \textsf{\textbf{B}} (along edge) as well as the corner \textsf{\textbf{C}}. \textbf{b,c} Corresponding d$I$/d$U$ spectra along those lines. At the corner \textsf{\textbf{C}}, indicated by a black solid arrow, we observe the strongest residual intensity at $-\Delta_\mathrm{tip}$ (dashed black line) corresponding to zero energy. }
        \label{fig:A3_STSlines}
    \end{figure}

\newpage
\subsection{Figure 8}
        \begin{figure}[!h]
        \centering
        \includegraphics[width=0.75\textwidth]{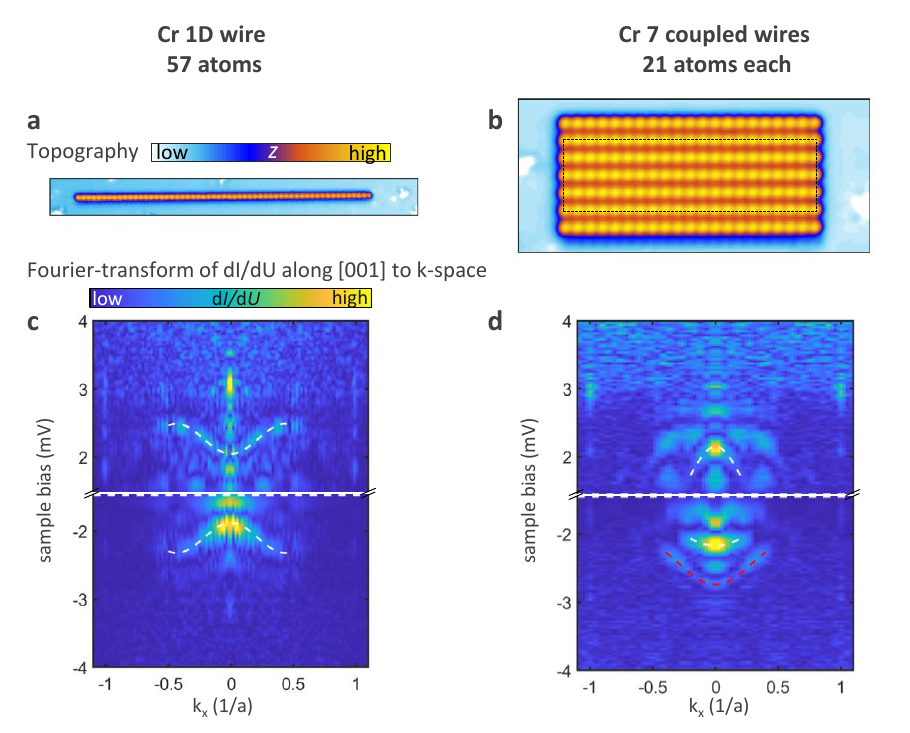}
        \caption{\textbf{Shiba bands along $[001]$ in momentum space.} Evolution of Shiba bands from 1D to 2D comparing a long 1D wire of Cr adatoms along the $[001]$ direction and \SI{0.66}{\nano\meter} inter-atomic distance to a two dimensional arrangement according to structure \textsf{A}. \textbf{a-b} Topography images.
        %\textbf{d-f} STS taken along the $[001]$ direction over each structure. For \textbf{f}, several lines through the bulk have been averaged indicated by the black dashed rectangle in \textbf{c}. Horizontal white dashed lines indicate the tip superconducting gap, vertical lines indicate the range used for Fourier transformation. 
        \textbf{c-d} FFT calculated from the STS data taken along the $[001]$ direction. For better visibility the plots are separated  to show the energy range $-4$mV...$-\Delta_\mathrm{tip}$ and $\Delta_\mathrm{tip}$...$4$mV with adjusted d$I$/d$U$ colorscale contrast. The momentum space plots suggest dispersing bands. The signal with highest intensity around $\pm2$mV is assigned to the $d_{z^2}$ atomic orbital (white dashed lines). Other bands arise from different single-atomic orbitals (red dashed line). From 1D to 2D, a change of sign in the dispersion relation is observed. This demonstrates the potential to engineer the band structure by advancement into two dimensions.
        %Once wires are coupled, an additional band becomes visible with onset around $\pm2.7$mV in \textbf{h-i}. The origin of this band is assigned to the $d_{yz}$ atomic orbital, whose extension along the $[1\overline{1}0]$ direction allows isolated $d_{yz}$ Shiba states to effectively hybridize once wires are coupled.
        }
        \label{fig:shibabandfft}
    \end{figure}

\begin{figure}
    \centering
    \includegraphics[width=\columnwidth]{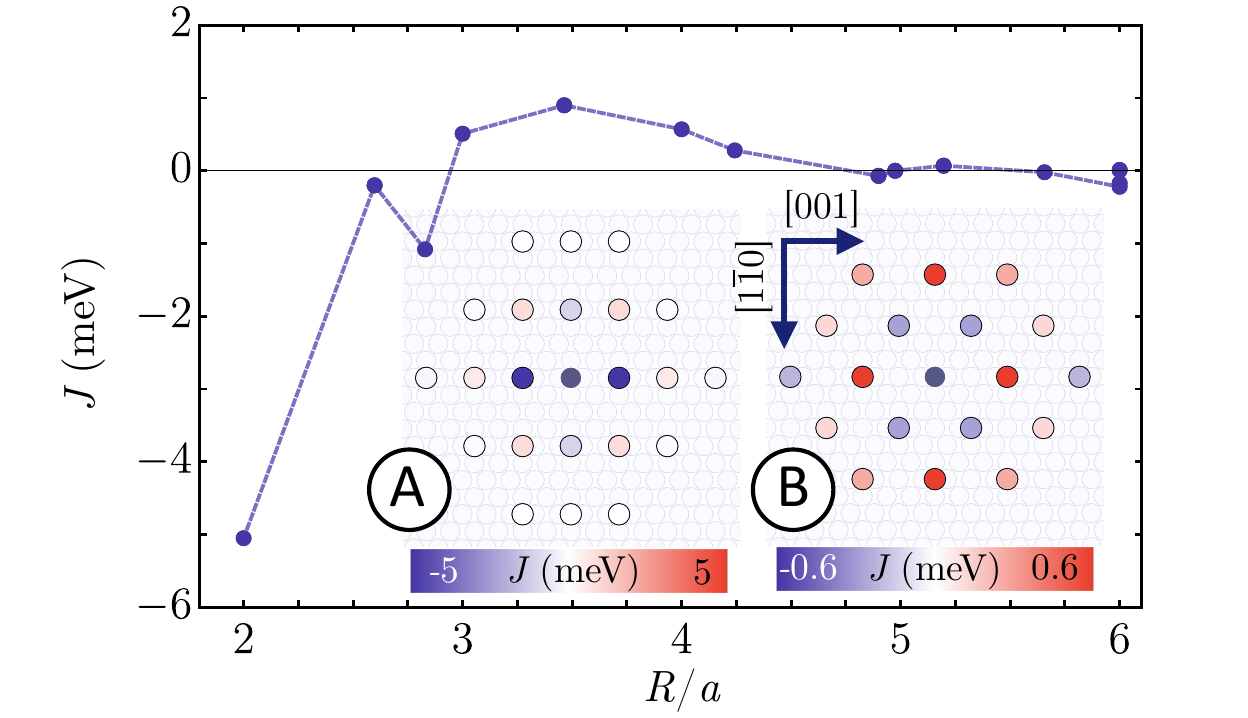}
    \caption{\textbf{Magnetic exchange interactions.}
Ab-initio Heisenberg exchange interactions $J$ as function of distance $R$ between Cr adatoms deposited on Nb(110), $a$ being the bulk Nb lattice parameter. Multiple data points at the same $R/a$ but with distinct values of $J$ correspond to symmetry non-equivalent sites. Positive (negative) values correspond to ferromagnetic (antiferromagnetic) coupling. The lower inset illustrates the simulated lattices, where each circle is colored as function of the size of $J$ with respect to the central atom (grey color). 
    }
    \label{fig:MEI}
\end{figure}
% \end{document}